\documentclass[paper]{JHEP3}
\pdfoutput=1
\usepackage{amsmath,amssymb,amsthm,amscd,graphicx,verbatim}
\input epsf.sty

\theoremstyle{definition}

\newcommand{\CA}{{\cal A}}
\newcommand{\CB}{{\cal B}}
\newcommand{\CC}{{\cal C}}

\newcommand{\CL}{{\cal L}}

\newcommand{\CO}{{\cal O}}

\newcommand{\CS}{{\cal S}}

\def\IZ{{\mathbb Z}}

\def\IP{{\mathbb P}}

\newcommand{\acycle}{{\cal A}}
\newcommand{\bcycle}{{\cal B}}

\newcommand{\tr}{{\rm Tr}}
\newcommand{\re}{{\rm e}}
\newcommand{\ri}{{\rm i}}
\newcommand{\rd}{{\rm d}}

\newcommand{\Ai}{\mathop{\rm Ai}\nolimits}

\newcommand{\be}{\begin{equation}}
\newcommand{\ee}{\end{equation}}
\newcommand{\ba}{\begin{aligned}}
\newcommand{\ea}{\end{aligned}}
\newcommand{\ben}{\begin{eqnarray}\displaystyle}
\newcommand{\een}{\end{eqnarray}}

\newcommand{\sectiono}[1]{\section{#1}\setcounter{equation}{0}}

\newcommand{\beq}{\begin{equation}} 
\newcommand{\eeq}[1]{\label{#1}\end{equation}} 
\newcommand{\bs}{\begin{split}} 
\newcommand{\es}{\end{split}}

\newdimen\tableauside\tableauside=1.0ex
\newdimen\tableaurule\tableaurule=0.4pt
\newdimen\tableaustep
\def\phantomhrule#1{\hbox{\vbox to0pt{\hrule height\tableaurule width#1\vss}}}
\def\phantomvrule#1{\vbox{\hbox to0pt{\vrule width\tableaurule height#1\hss}}}
\def\sqr{\vbox{%
  \phantomhrule\tableaustep
  \hbox{\phantomvrule\tableaustep\kern\tableaustep\phantomvrule\tableaustep}%
  \hbox{\vbox{\phantomhrule\tableauside}\kern-\tableaurule}}}
\def\squares#1{\hbox{\count0=#1\noindent\loop\sqr
  \advance\count0 by-1 \ifnum\count0>0\repeat}}
\def\tableau#1{\vcenter{\offinterlineskip
  \tableaustep=\tableauside\advance\tableaustep by-\tableaurule
  \kern\normallineskip\hbox
    {\kern\normallineskip\vbox
      {\gettableau#1 0 }%
     \kern\normallineskip\kern\tableaurule}%
  \kern\normallineskip\kern\tableaurule}}
\def\gettableau#1{\ifnum#1=0\let\next=\null\else
\squares{#1}\let\next=\gettableau\fi\next}

\newcommand{\figref}[1]{Fig.~\protect\ref{#1}}

\title{The topological open string wavefunction}

\author{
Alba Grassi, Johan K\"all\'en and Marcos Mari\~no
\\
D\'epartement de Physique Th\'eorique et Section de Math\'ematiques,\\
Universit\'e de Gen\`eve, Gen\`eve, CH-1211 Switzerland\\
\\
\email{alba.grassi@unige.ch}, \quad
\email{johan.kallen@unige.ch}, \quad
\email{marcos.marino@unige.ch}
}

\abstract{We show that, in local Calabi--Yau manifolds, the topological open string partition function transforms as a wavefunction under modular transformations. Our derivation is based on the topological recursion for matrix models, and it generalizes in a natural way the known result for the closed topological string sector. As an application, we derive results for vevs of $1/2$ BPS Wilson loops in ABJM theory at all genera in a strong coupling expansion, for various representations.}

\begin{document}

\sectiono{Introduction}
Topological string theory on Calabi--Yau (CY) manifolds has been an important source of results in string theory, gauge theory and mathematics (see for example \cite{mmhouches, mmreview, mmbook, vonk, nv-review} for reviews). In the so-called local case, where the CY is non-compact, the theory can be solved exactly, by using for example large $N$ techniques in matrix models \cite{dv,mmopen,bkmp} or the theory of the topological vertex \cite{akmv-tv}. 

Closed topological string amplitudes satisfy many interesting properties. In the local case, and from the B-model point of view, they can be regarded as  holomorphic objects associated to an algebraic curve or Riemann surface. They depend on a choice of ``symplectic frame", i.e. on a choice of symplectic basis for the homology of the Riemann surface, and they turn out to have non-trivial transformation properties under a change of basis or modular transformation. Equivalently, one can introduce a non-holomorphic dependence in the amplitudes which is governed by the holomorphic anomaly equations of \cite{bcov}. As shown in \cite{abk}, the transformation properties of the closed string amplitudes can be derived from the fact that the total closed string partition function (summed over all genera) is a wavefunction \cite{witten}. Modular transformations correspond to canonical transformations, which lift quantum-mechanically to integral transforms of the wavefunction. Therefore, a change of symplectic basis leads to an integral transform of the topological closed string partition function. 

These properties of the closed topological string amplitudes can be also derived by using the solution of the B-model in terms of matrix integrals \cite{dv,mmopen,bkmp}. This solution is based on the topological recursion of Eynard and Orantin \cite{eo}, which encodes as well the modular properties of the resulting amplitudes. It was shown in \cite{emo} that the modular behavior of the closed string amplitudes, as deduced from the topological recursion, agrees indeed with the wavefunction behavior of the partition function found in \cite{witten,abk}. 

The topological recursion of \cite{eo} gives as well a method to compute the modular transformation of {\it open} string amplitudes. In this paper, we show that these properties can be summarized by saying that the total open string partition function transforms as a wavefunction. This generalizes the results of \cite{witten,abk} to the open sector, since the closed string partition function can be regarded as a specialization of the open string partition function where all the open moduli are set to zero. 

The wavefunction behavior of the open string partition function has practical applications, since it makes possible to 
relate in a precise way open string amplitudes in different frames. 
One interesting situation where this can be used is the calculation of vacuum expectation values (vevs) of $1/2$ BPS Wilson loops \cite{dt} in 
ABJM theory \cite{abjm}. These vevs can be computed by localization, 
in terms of a matrix model \cite{kwy,dt}. It turns out that they are given by open topological string amplitudes in a non-compact CY, local 
$\IP^1 \times \IP^1$ \cite{mp}, but in the so-called {\it orbifold} frame \cite{akmv}. 
As an application of the main result of this paper, we obtain results for the vevs of $1/2$ BPS Wilson loops by first performing the calculation 
in the large radius frame, and then using the fact that the open string partition function is a wavefunction. We find in this way 
all-genus results for vevs of $1/2$ BPS Wilson loops as integral transforms of topological string amplitudes at large radius. These expressions are exact in 
$k$, the coupling of ABJM theory, but they are expanded around the strong coupling limit. They correspond to the M-theory expansion of the amplitudes discussed in for 
example \cite{mp-fermi,kmss}. In particular, we rederive in this way the 
result for $1/2$ BPS Wilson loop vev with winding $n$ derived in \cite{kmss} in the M-theory regime, and we extend it to other representations. 
Our method also makes it possible to calculate systematically worldsheet instanton corrections, which are difficult to obtain in the Fermi gas approach of \cite{kmss}.

The wavefunction behavior of the open string amplitudes has been addressed before. In \cite{adkmv,anv,kp}, 
the behavior of the open string partition function has been studied as one changes the {\it open} moduli, although 
as far as we know there is no general statement for this behavior. The paper \cite{nw} studies the wavefunction behavior of the 
open string partition function in the compact CY case. 
 
This paper is organized as follows. In section 2 we review the definition and construction of topological open string amplitudes and the topological 
recursion of Eynard and Orantin. In section 3 we derive our main result, namely, we show that the total, topological open string partition function, 
transforms as a wavefunction under modular transformations. In section 
4 we use our main result to obtain expressions for $1/2$ BPS Wilson loop vevs at all orders in the genus expansion and expanded at strong coupling. 
Finally, in section 5 we end up with some conclusions and prospects for future work. 

\sectiono{Open topological string amplitudes and topological recursion}

\subsection{Open topological string amplitudes}

In this paper we will study open topological string amplitudes in local CY geometries. There are two types of local CYs which are particularly interesting. The first ones are of the form 
\be
\label{cydv}
uv=H(x, y), \quad H(x,y)=y^2-(W'(x))^2+f(x),
\ee
where $W(x), f(x)$ are polynomials of degree $d+1$, $d-1$ respectively. The Riemann surface $H(x,y)=0$ associated to this geometry is 
the hyperelliptic curve 
\be
\label{hypercurve}
y^2= (W'(x))^2-f(x),
\ee
of genus $n=d-1$. Dijkgraaf and Vafa conjectured in \cite{dv} that type B topological string theory on these backgrounds 
is equivalent to a matrix model with potential $W(x)$ and $d-1$ cuts  (see \cite{mmhouches} for a review). 

A more interesting class of local geometries are toric CY manifolds, which are non-compact. In this case, both open and closed topological string amplitudes have an enumerative meaning in the A-model, which we now review briefly (see for example \cite{mmreview} for a presentation with appropriate references). 
Closed string amplitudes at genus $g$ can be expressed as a sum over instanton sectors. These are labelled by a class $\beta \in H_2(X)$, where $X$ is the CY target, and they read
\be
F_g(t)= \sum_{\beta \in H_2(X)} N_{g, \beta}\,  \re^{-\beta\cdot t}.
\ee
In this equation, $t$ denotes the vector of closed K\"ahler moduli. The rational numbers $N_{g, \beta}$ are Gromov--Witten invariants counting holomorphic maps from a Riemann surface of genus $g$, $\Sigma_g$, to the CY $X$, and in the class $\beta$. It is useful to define the total closed string 
free energy as:
\begin{equation}
F(g_s,t)=\sum_{g=0}^{\infty} F_g(t) g_s^{2g-2}.
\label{freen}
\end{equation}
Gopakumar and Vafa \cite{gv} showed that the generating
functional (\ref{freen}) can be written as a generalized index that
counts BPS states in M-theory compactified
on $X$, and this leads to the following structural result for
$F(g_s, t)$:
\begin{equation}
F(g_s,t)=\sum_{g=0}^{\infty} \sum_{\beta} \sum_{m=1}^{\infty}
n_{g,\beta} {1\over m}\biggl( 2 \sinh {m g_s \over 2} \biggr)^{2g-2} \re^{-m\beta\cdot t},
\label{gvseries}
\end{equation}
where $n_{g,\beta}$, known as Gopakumar-Vafa invariants, are {\it integer}
numbers. 

In order to define open topological strings on a CY $X$, we need to specify boundary conditions. This is done by choosing a brane wrapping 
a Lagrangian submanifold  $\CL \subset X$. The free energy of the open topological string theory can be obtained by summing
the contribution of open worldsheet string instantons in different topological sectors. These sectors classify maps from an open Riemann surface $\Sigma_{g,h}$ to $X$, in such a way that the 
boundaries of $\Sigma_{g,h}$ are mapped to $\CL$. They are labelled by two different kinds of data:
the boundary part and the bulk part. The bulk part is labelled by relative homology classes 
$\beta \in H_2(X, \CL)$. We will assume that
$b_1 (\CL)=1$ (as it happens for the Lagrangian submanifolds constructed in \cite{av}) 
so that $H_1 (\CL)$ is generated by one nontrivial
one-cycle. Then, the topological sector of the boundary is classified by winding numbers $\ell_i$ specifying how many times the 
boundaries of $\Sigma_{g,h}$ wrap the non-trivial one-cycle of $\CL$. We will collect these integers into a single
$h$-tuple denoted by ${\boldsymbol \ell}=(\ell_1, \cdots, \ell_h)$.

There are various amplitudes that we can consider, depending on
the topological data that we want to keep fixed. It is very useful to fix
$g$ and the winding numbers, and sum over all bulk classes. This produces
the following generating
functional of open Gromov-Witten
invariants:
\begin{equation}
F_{g, \boldsymbol \ell} (t) =\sum_{\beta} N_{g,\beta, {\boldsymbol \ell}}\,\re^{-\beta\cdot t}.
\end{equation}
In this equation, the sum is over relative homology classes $\beta \in H_2(X,\CL)$. The quantities $N_{g, \beta, {\boldsymbol \ell}}$
are open Gromov-Witten invariants.
They ``count" in an appropriate sense the number of holomorphically
embedded Riemann surfaces of genus $g$ in $X$ with Lagrangian boundary
conditions specified by $\CL$, and in the class represented
by $\beta, {\boldsymbol \ell}$. They are in general rational numbers.

In order to consider all topological sectors, we have to introduce a $U(\infty)$ matrix $V$ which takes into account
different sets of winding numbers ${\boldsymbol \ell}$. The total open topological string free energy is defined by
\be
F(V)=  \sum_{g=0}^{\infty} \sum_{h=1}^{\infty}
\sum_{{\boldsymbol \ell}} {1 \over h!}
g_s^{2g-2+h} F_{g,{\boldsymbol \ell}} (t)
{\rm Tr}\,V^{\ell_1} \cdots {\rm Tr}\, V^{\ell_h}.
\label{totalfreeopen}
\ee
Open topological string amplitudes have an integrality structure 
discovered in \cite{ov,lmv}. It turns out that the total free energy can be written as 
\be
\begin{aligned}
 F(V)=& \sum_{\beta \in H_2(X, L)} \sum_{g=0}^\infty \sum_{h=1}^\infty \sum_{{\boldsymbol \ell}} \sum_{m=1}^\infty 
 {1\over h!} n_{g, \beta, {\boldsymbol \ell}} {1\over m} \left( 2\sinh {m g_s \over 2} \right)^{2g-2} \\
& \qquad \qquad \qquad \cdot  \prod_{i=1}^h \left(2 \sinh {m \ell_i g_s \over 2}  \right) {1\over \ell_1 \cdots \ell_h}  \tr\, V^{m \ell_1} \cdots \tr \, V^{m \ell_h}{\rm e}^{-m\beta \cdot t}.
\end{aligned}
\label{multopen}
\ee
In this expression, $n_{g, \beta, {\boldsymbol \ell}}$ are integer invariants which 
generalize the Gopakumar--Vafa invariants of closed topological strings 
(in fact, as shown in \cite{lmv}, the invariants $n_{g, \beta, {\boldsymbol \ell}}$ can 
be written in terms of a more fundamental set of integer invariants, but we will not need them in this paper). 

We will often write the free energy as 
\be
F(V)=\sum_R W_R \tr_R\, V,
\ee
where the sum is over $U(\infty)$ representations, while the total open string partition function is defined as 
\be
Z(V)=Z_{\rm cl} \exp(F(V)).
\ee
Here, we used the total closed string free energy (\ref{freen}) to define the closed string partition function, 
\be
Z_{\rm cl}=\exp\left(F(g_s,t)\right).
\ee
We will write $Z(V)$ sometimes as 
\be
\label{zr}
Z(V)  =\sum_R Z_R \tr_R \, V. 
\ee

It was conjectured in \cite{mmopen,bkmp} that type B topological string theory on mirror of toric CY manifolds can be solved in terms of the topological 
recursion of \cite{eo} (this conjecture has been recently proved in \cite{eop}). Since this formalism describes as well the solution to the $1/N$ 
expansion of matrix models, this generalizes the conjecture of \cite{dv} to backgrounds with 
an enumerative meaning. We can then use the formalism of topological recursion to provide a 
unified description of open and closed topological string amplitudes in local CY geometries. 

\subsection{Open strings and topological recursion} \label{toprec}

The formalism of topological recursion of \cite{eo} starts with an algebraic curve $H(x,y)=0$ of genus $\bar g$. 
We will choose a canonical basis of cycles on it:
\be
\underline{\acycle}_I \cap \underline{\bcycle}_J=\delta_{IJ}, 
\qquad 
\underline{\acycle}_I \cap\underline{\acycle}_J=0, 
\qquad
 \underline{\bcycle}_I \cap \underline{\bcycle}_J=0, \qquad  I,J = 1, \cdots, \bar g.
\ee
There are $\bar g$ linearly independent holomorphic forms $\omega_I$ on $H(x,y)$ normalized on the $\underline{\acycle}$-cycles:
\be
 \oint_{\underline{\acycle}_I}  \omega_J = \delta_{IJ}, \qquad  I,J = 1, \cdots, \bar g,
\ee
and the Riemann matrix of periods, $\tau$, is a symmetric $\bar g \times \bar g$ matrix defined by
\be
\oint_{ \underline{\bcycle}_J} \omega_I = \tau_{IJ}.
\ee
On the curve $H(x,y)=0$ there exists a unique bilinear form $\underline{B}(p,q)$ with a unique double pole
at $p=q$ without residue, and normalized on the $\underline{\acycle}$ cycles:
\be
\underline{B}(p,q) \mathop{\sim}_{p \to q} {\rd z(p) \rd z(q) \over (z(p)-z(q))^2} \;  + \; \hbox{finite}, \qquad 
\oint_{\underline{\CA}}\underline{B} = 0.
\ee
Here, $z$ is any local parameter on the curve, and $\underline{B}$ is usually called the {\it Bergmann kernel}. The Bergmann kernel has the following properties  \cite{eo}
\be
\underline{B}(p,q)=\underline{B}(q,p), \quad \oint_{\underline{\CB}_I}\underline{B}=2\pi \ri \omega_I ~.
\ee

Following the procedure in \cite{eo,emo}, we will now introduce a new set of cycles, $ \CA,  \bcycle$, 
depending on an arbitrary complex symmetric matrix $\kappa$:
\be
\label{kappa-intro}
 \bcycle :=\underline \CB - \tau \underline \CA, \qquad 
 \CA :=\underline\CA - \kappa \CB.
\ee
We then define a $\kappa$--modified Bergmann kernel ${B}$, normalized on these new
 cycles, and satisfying thus
\be
{B}(p,q)\mathop{\sim}_{p \to q} {\rd z(p) \rd z(q) \over (z(p)-z(q))^2} \;  + \; \hbox{finite}, \qquad 
\oint_{ \CA} {B} = 0.
\ee
This definition implies the relation
\be
 {B}(p,q)= \underline{B}(p,q) + 2 \pi \ri\sum_{I,J} \omega_I(p)\, \kappa^{IJ }\, \omega_J(q).
\ee
Notice that for the new $\bcycle$-cycles we have the following relations:
\be \label{bcycleBerg}
\oint_{\bcycle_I}B=2\pi \ri \omega_I~,\quad \oint_{\bcycle_I}\omega_J=0~.
\ee
With these ingredients, one defines recursively an infinite set of symmetric meromorphic differentials $W_h^{(g)}$ on the curve, as follows. Let $a_i$ be the branching points of the curve. If $q$ is near a branchpoint, there is by definition a unique point $\overline q$ such that $x(q)=x(\overline q)$. The starting point of the recursion is 
\be
\ba
W_h^{(g)}&=0 \quad {\rm if}\,\, g<0, \\
W_1^{(0)}(p)& = 0, \\
W_2^{(0)}(p_1,p_2)& = {B}(p_1,p_2).
\ea 
\ee
The recursion is then given by 
\be\label{defWkgrecursive}
\ba
& W_{h+1}^{(g)}(p,p_1,\dots,p_h) \\
&=\sum_i {\rm Res}_{q= a_i} {\rd E_{q}(p)\over \omega(q)}\,\left(\sum_{m=0}^g \sum_{J\subset H} W_{|J|+1}^{(m)}(q,p_J)W_{h-|J|+1}^{(g-m)}(\overline{q},p_{H/J})
+ W_{h+2}^{(g-1)}(q,\overline{q},p_H) \right).
\ea
\ee
Notice that it follows that all $W^{(g)}_h$'s have vanishing ${\CA}$-cycle integrals. In this equation, $q$ is taken to be near a branchpoint, and
\be
\omega(q) = (y(q)-y(\overline{q}))\rd x(q), \qquad 
\rd E_{q}(p) = {1\over 2} \int_{q}^{\overline{q}} {B}(\xi,p)
\ee
where the integration path lies entirely in a vicinity of $a_i$. 
If $J=\{ i_1,i_2,\dots,i_j\}$ is a set of indices, we write $p_J=\{ p_{i_1},p_{i_2},\dots,p_{i_j}\}$. In the equation above we have $H=\{1,2,\dots,h\}$, and the summation over $J$ is over all subsets of $H$.

Once these differentials are constructed, one can compute the closed string free energies $F^{(g)}$ for $g\ge 2$ as 
\be
F^{(g)} = {1\over 2g-2}\,\sum_i {\rm Res}_{q= a_i} \Phi(q) W_{1}^{(g)}(q),
\ee
where $\Phi(q) = \int^q \lambda$ is any antiderivative of the meromorphic differential
\be
\lambda =y \rd x,
\ee
which satisfies 
\be
\partial_I \lambda= (2\pi \ri)^{\frac{1}{2}} \omega_I.
\ee

The meromorphic differentials $W^{(g)}_h$ defined by the topological recursion 
are functions of two types of variables. On the one hand, we have the {\it open string moduli}, which are the variables $p_i$ 
upon which they depend. On the other hand, they depend on the {\it closed string moduli}, 
which are the complex moduli of the spectral curve itself. These closed string moduli can be parametrized by the 
$\underline{\CA}$-periods of $\lambda$ 
\be
t^I ={1\over( 2 \pi \ri)^{1/2}}  \oint\limits_{\underline\CA_I}{\lambda}. 
\ee

In this formalism, both $F^{(g)}$ and the forms $W_h^{(g)}$ depend as well on the matrix-valued parameter $\kappa$ which we have introduced in (\ref{kappa-intro}). 
The usual topological string or matrix model amplitudes are obtained by setting $\kappa=0$ in the above formalism, i.e. 
by using the topological recursion but with the standard Bergmann kernel. To recover the standard open string amplitudes as defined for example in (\ref{totalfreeopen}), we define the 
integrated forms $A^g_h$, $h> 0$ by
\beq
A^{(g)}_h (t,\kappa) = \int {W^{(g)}_h (t,\kappa)}~,
\eeq{wtoa}
except for $(g,h)=(0,1)$ and $(g,h)=(0,2)$. In those cases we have
\beq
A^{(0)}_1=-\int{\lambda},
\eeq{a01}
and
\beq
A^{(0)}_2= \int{\left( B(p_1,p_2)-\frac{\rd p_1 \rd p_2}{(p_1-p_2)^2}\right)}.
\eeq{a02}
The differentials $W^{(g)}_h$ have an expansion in inverse powers of the open string moduli $p_i$, and the integrated amplitudes have then an expansion of the form 
\be
\label{a-ah}
A^{(g)}_h (t, \kappa,z_1, \cdots, z_h)=\sum_{{\boldsymbol \ell}} A_{{\boldsymbol \ell}} ^{(g)}(t, \kappa) z_1^{\ell_1} \cdots z_h^{\ell_h},
\ee
where $z_i=p_i^{-1}$ and, as above, $\boldsymbol \ell=(\ell_1, \cdots, \ell_h)$. The coefficients of this expansion, evaluated at $\kappa=0$, are the topological open string amplitudes for boundary conditions specified by $\boldsymbol \ell$:
\be
F_{g,{\boldsymbol \ell}}(t)=A_{{\boldsymbol \ell}}^{(g)}(t, \kappa=0).
\ee

\sectiono{The topological open string partition function as a wavefunction}

\subsection{Symplectic transformations} \label{sec:symptransf}

The construction of the open and closed string amplitudes through the topological recursion depends on a choice of symplectic frame, i.e. on a choice of a distinguished set of $\underline \CA$, $\underline \CB$ cycles on the curve $H(x,y)=0$. 
A natural and important question in the study of topological string theory and matrix models is: how do $F^{(g)}$ and $W_h^{(g)}$ change under a change of symplectic frame? We will refer to these transformations 
of the amplitudes as one changes the symplectic frame as {\it symplectic} or {\it modular} transformations. 
In the case of the closed string free energies $F^{(g)}$, a detailed answer was obtained in \cite{abk} by using the fact that, as pointed out in \cite{witten}, 
the total closed string partition function can be regarded as a wavefunction. This was based in turn on the holomorphic anomaly equations of \cite{bcov}. 

We recall that a modular or symplectic transformation for a curve $H(x,y)=0$ of genus $\bar g$ is implemented by a symplectic matrix 
\be
\Gamma=\begin{pmatrix}A& B \\
               C&D \end{pmatrix}  \in {\rm Sp}(2\bar g, \IZ)
\label{simp}
\ee
where the $\bar g \times \bar g$ matrices $A$, $B$, $C$, $D$, with integer-valued entries, satisfy
\be
A^{\rm T}D-C^{\rm T}B= {\bf 1}_{\bar g}, \,\,\,\ 
A^{\rm T}C=C^{\rm T}A, \,\,\,\ 
B^{\rm T}D=D^{\rm T}B.
\label{simple}
\ee
The cycles of the curve change as 
\be\label{Cycletransform}\left( \begin{array}{c}
\underline{\CB} \\ 
\underline{\CA}
\end{array} \right) \rightarrow \left( \begin{array}{cc}
A & B \\ 
C & D
\end{array} \right) \left( \begin{array}{c}
\underline{\CB }\\ 
\underline{\CA}
\end{array} \right), \ee
while the period matrix $\tau$ of the curve changes as 
\be
\tau \rightarrow  (A\tau+B)(C\tau + D)^{-1}.
\ee

The formalism of \cite{eo} reviewed in the previous section gives a direct way of deriving the modular properties of the amplitudes, 
through the incorporation of the $\kappa$ parameter. In fact, there are two equivalent ways of 
understanding these properties, as emphasized in \cite{abk}. In the first point of view, one considers the topological string amplitudes, which are the holomorphic objects $F^{(g)}(t,0)$ and $W_h^{(g)}(t,0)$ for $\kappa=0$. Then, under a modular transformation implemented 
by $\Gamma$, the amplitudes change as follows, 
\be
\label{mod-trans}
\ba
W^{(g)}_h(t,0) \rightarrow W^{(g)}_h (t,\kappa), \\
F^{(g)} (t,0) \rightarrow F^{(g)} (t,\kappa), 
\ea
\ee
where 
\be
 \kappa=-(\tau+C^{-1}D)^{-1}.
 \ee
One important fact of these transformation properties is that the open string moduli do not change under a modular transformation. However, open string amplitudes evaluated in different regions of moduli space require 
different parametrizations of these moduli, and one needs to redefine them by an overall factor which depends on the closed string moduli, as first found in \cite{akv}. In general, one has
\be
\label{open-m-m}
p_i \rightarrow p_i \exp \left[ \sum_I 
\left( a_I t_I + b_I t_I^{\rm bare} \right)\right],
\ee
where $t_I^{\rm bare}$ are the ``bare" closed string moduli, corresponding to complex deformation parameters of the spectral curve, 
and $a_I$, $b_I$ are rational numbers which can be found by a detailed analysis of the geometry, 
see for example \cite{bkmp} for a detailed explanation. This is often called the 
open string mirror map, or the choice of open flat coordinate. There is then a canonical choice of open moduli, given by the solution of the topological recursion, and other choices can be obtained by using (\ref{open-m-m}). 
 
As shown in \cite{emo}, the transformation of the closed string amplitudes in (\ref{mod-trans}) is equivalent to the statement of \cite{witten,abk} that the closed string partition function transforms as a wavefunction. More precisely, it was shown in \cite{emo} that the total $\kappa$-dependent partition function 
\be
 \ln{(Z(t,\kappa))}=F(t, \kappa) = \sum_{g=0}^{\infty} g_s^{2-2g} F^{(g)}(t, \kappa)
\ee
can be obtained as an integral transform of the partition function with $\kappa=0$, $Z(\eta, 0)= \exp F(\eta, \kappa=0)$, 
\be
\label{genfunanti}
Z(t,\kappa)=  \int \rd \eta\, \re^{-S(t, \eta, \kappa) g_s^{-2}+ F(\eta,0)},
\ee
where 
\be
\label{noyau}
S(t, \eta, \kappa)= {1\over 2}  (\eta-t) \kappa^{-1} (\eta-t)+ (\eta -t)^I \partial_I F^{(0)}(t,0 )+ 
{1\over 2}  (\eta-t)^I \partial^2 _{IJ}  F^{(0)}(t,0) (\eta-t)^J. 
\ee
The integral transform is evaluated in a genus expansion, by doing a saddle-point evaluation of the integral for small $g_s$. 

In the second point of view on modular transformations, one considers the amplitudes $W^g_h(t,\kappa)$, $F^{(g)}(t, \kappa)$ with 
\be
\label{ka-nh}
\kappa=-(\tau-\overline \tau)^{-1}.
\ee
In this case, the resulting amplitudes are modular invariant, as shown in \cite{eo}, but they inherit a non-holomorphic dependence through the conjugate 
$\overline \tau$ appearing in (\ref{ka-nh}). It was shown in \cite{emo} that, for the choice of $\kappa $ in (\ref{ka-nh}), 
the closed string amplitudes $F^{(g)}(t, \kappa)$ satisfy the holomorphic anomaly equations of \cite{bcov}, specialized to local geometries \cite{kz}. 

The purpose of this paper is to generalize to the open string sector the results of \cite{emo} concerning the wavefunction behavior of the closed string partition function. As we will illustrate in a moment, the transformations (\ref{mod-trans}) are quite complicated when written down for the individual amplitudes. It is a non-trivial fact that, when we organize the amplitudes in terms of partition functions, these transformation properties can be elegantly summarized by a wavefunction behavior, i.e. by an integral transform of the partition function. 

\subsection{The wavefunction behavior}

\begin{figure}
\centerline{\includegraphics[width=0.8\linewidth,angle=0]{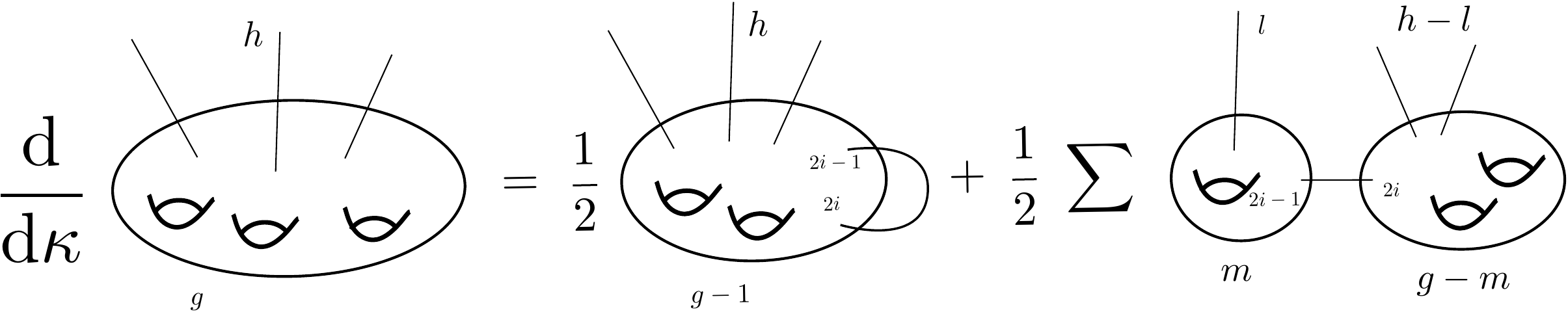} }
\caption{ A graphic representation of the equation (\ref{variatW}).}
 \label{Fig:recursion}
 \end{figure}

In order to show that the total open topological string partition function transforms as a wavefunction, one has to be more explicit about the transformations (\ref{mod-trans}), i.e. one should compute $W^{(g)}_h(t,\kappa)$ and $F^{(g)}(t,\kappa)$ in terms of $W^{(g)}_h(t,0)$ and $F^{(g)}_h(t,0)$. As explained in \cite{emo}, the basic observation is that the $\kappa$ dependence of the $W_h^{(g)}$'s enters only through the Bergmann kernel, therefore each $W_h^{(g)}$ is a polynomial in $\kappa$ of degree at most $3g-3+2h$: 
\be
W^{(g)}_h(t,\kappa) = \sum_{m=0}^{3g-3+2h}\frac{\kappa^m}{m!} {\rd^m W^{(g)}_h\over \rd\kappa ^m}(t,0).
\eeq{Wexpkappa}
In order to obtain this polynomial, it is convenient to compute $\rd W_h^{(g)}/\rd \kappa$. This was done in \cite{eo} and the result is:
\be
\label{variatW}
\ba
2\pi \ri {\partial \over \partial \kappa_{IJ}}\,W^{(g)}_{h}(p_H)
&=  {1\over 2}\,\oint_{r\in\bcycle_J}\oint_{s\in\bcycle_I} W^{(g-1)}_{h+2}(p_H,r,s) \\
& + {1\over 2}\,\sum_{m=0}^g \sum_{L\subset H} \oint_{r\in\bcycle_I} W^{(m)}_{|L|+1}(p_L,r) \oint_{s\in\bcycle_J} W^{(g-m)}_{h-|L|+1}(p_{H/L},s). 
\ea
\ee
In particular, for $h=0$, 
\be\label{variatF}
 2\pi \ri{\partial \over \partial \kappa_{IJ}}\,F^{(g)}
=  {1\over 2}\,\oint_{r\in\bcycle_J}\oint_{s\in\bcycle_I} W^{(g-1)}_{2}(r,s) + {1\over 2}\,\sum_{m=1}^{g-1} \oint_{r\in\bcycle_I} W^{(m)}_{1}(r) \oint_{s\in\bcycle_J} W^{(g-m)}_{1}(s), 
\qquad g\geq 2.
\ee
The recursion relations (\ref{variatW}) and (\ref{variatF}) can be written in terms of diagrams \cite{emo} where the $W^{(g)}_h$'s and $F^{(g)}$'s are represented by  Riemann surfaces with $g$ holes and $h$ legs sticking out. The integrals over $\mathcal{B}$ cycles  are represented by legs which start and end on Riemann surfaces. This is illustrated in  \figref{Fig:recursion}.

Notice that the equations (\ref{variatW}) have the same structure that the topological recursion itself. They can be iterated to calculate the expansion (\ref{Wexpkappa}),  in terms of integrals of $W^{(g)}_h(t,0)$ around $\CB$-cycles. For example, one finds, for $W^{(1)}_1(t,\kappa)$, 
\be \ba\label{totest}W^{(1)}_1(t,\kappa) &=
W^{(1)}_1(t,0)+\frac{\kappa^{IJ}}{2\pi \ri} \left({1\over 2} \oint\limits_{\mathcal{B}_I}
\oint\limits_{\mathcal{B}_J}{W^{(0)}_3} +\oint\limits_{\mathcal{B}_I}W^{(1)}_1\oint\limits_{\mathcal{B}_J}{W^{(0)}_2}\right)
\\
&+\frac{\kappa^{IJ}\kappa^{MN}}{(2\pi \ri)^2}\left({1\over 2} \oint\limits_{\mathcal{B}_I}\oint\limits_{\mathcal{B}_J} \oint\limits_{\mathcal{B}_M}{W^{(0)}_3} \oint\limits_{\mathcal{B}_N}W^{(0)}_2 \right)~,
 \ea\ee
where we have used $\eqref{bcycleBerg}$ to show that 
 \be
 \oint_{\bcycle_I}\oint_{\bcycle_J} W^{(0)}_2=0~.
 \ee
It is illuminating to verify this transformation law in the case of an elliptic curve by using the explicit expressions for the $W^{(g)}_h$. This we do in the Appendix. 

After using the recursion relations we end up with integrals of the form
\beq
\underbrace{\oint\limits_{\mathcal{B}}\dotsb\oint\limits_{\mathcal{B}}}_{n}{W^{(g)}_h}~,
\eeq{}
where $h\geq n$. We can rewrite them as derivatives with respect to $t$. We have that \cite{eo} 
\begin{align}
\underbrace{\oint\limits_{\mathcal{B}}\dotsb\oint\limits_{\mathcal{B}}}_{n}{W^{(g)}_h}&=(-1)^n\partial^n W^{(g)}_{h-n}~, \quad \quad h>n+1, \,\, g\geq 0~, \label{intder1} \\
\underbrace{\oint\limits_{\mathcal{B}}\dotsb\oint\limits_{\mathcal{B}}}_{n}{W^{(g)}_h}&=(-1)^n\partial^n W^{(g)}_1~, \quad \quad h=n+1,\,\, g>0~, \label{intder2}\\
\underbrace{\oint\limits_{\mathcal{B}}\dotsb\oint\limits_{\mathcal{B}}}_{n}{W^{(0)}_h}&=2\pi \ri~(-1)^{n-1}~ \partial^{n-1} \omega~, \quad \quad h=n+1, \,\, g=0~, \label{intder3}\\
\underbrace{\oint\limits_{\mathcal{B}}\dotsb\oint\limits_{\mathcal{B}}}_{n}{W^{(g)}_h}&=(-1)^{n}\partial^n F^{(g)}~, \quad \quad h=n, \,\,  g\geq0~, \label{intder4}
\end{align}
where the derivatives are w.r.t.\ 
\be
\epsilon=(2\pi \ri)^{-\frac{1}{2}}t
\ee
and $\eqref{intder4}$ does not hold for $(g,h)=(0,1),(0,2)$. As can be seen from the formulas in section \ref{toprec}, in these two cases we get zero on the r.h.s. in $\eqref{intder4}$. 

\begin{figure}
\centerline{\includegraphics[width=0.65\linewidth,angle=0]{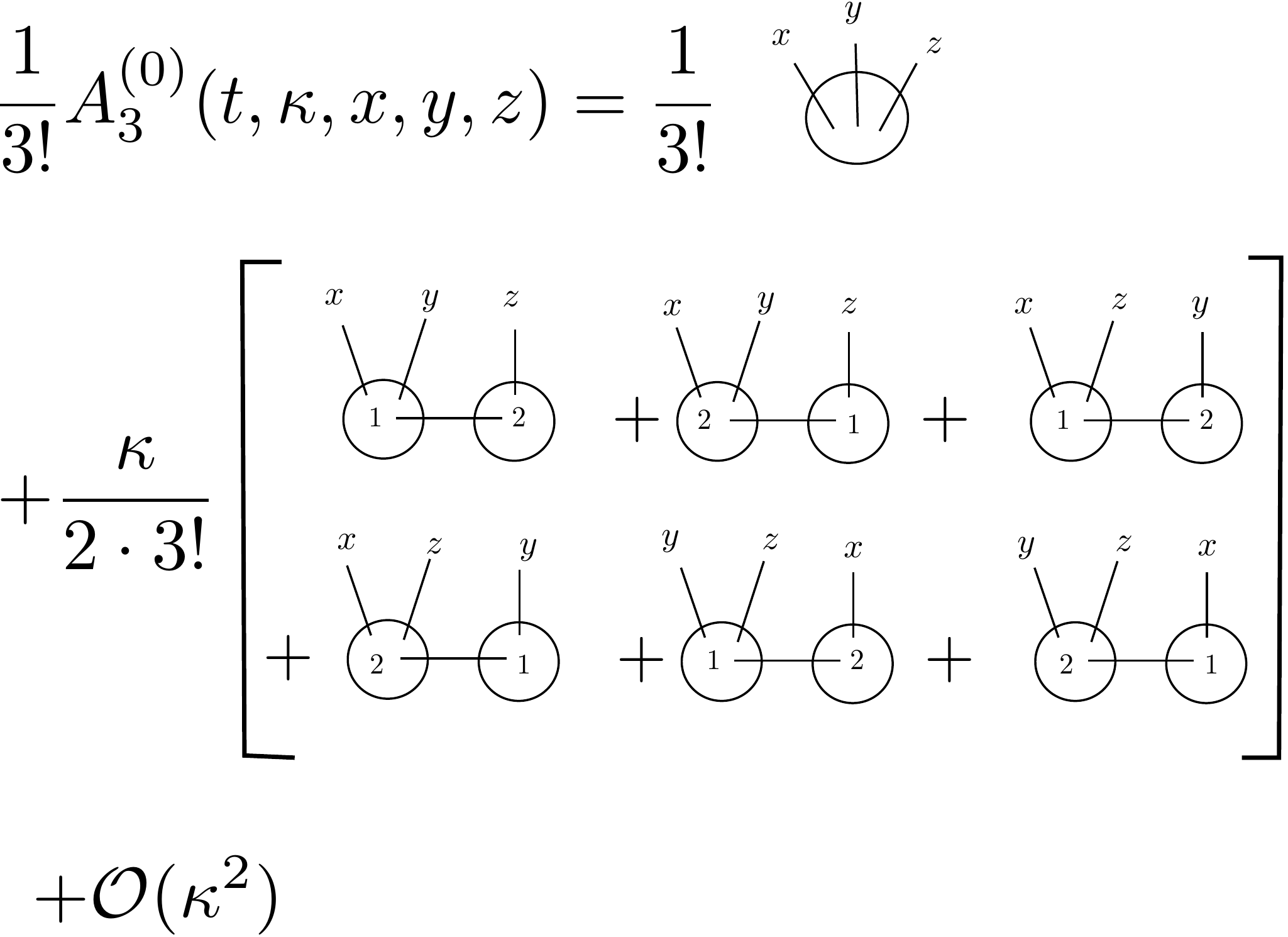} }
\caption{ Graphs that contribute to $A^{(0)}_3(t,\kappa,x,y,z)$ after iterating (\ref{variatW}), up to order $\kappa$.  }
 \label{graphproof}
 \end{figure}

\begin{figure}
\centerline{\includegraphics[width=0.8\linewidth,angle=0]{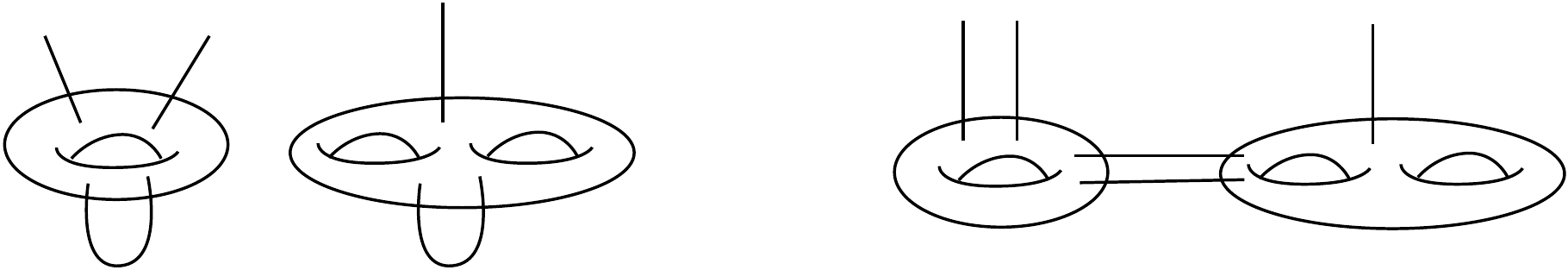} }
\caption{ A graphic representation of a disconnected (left) and connected (right) surface.}
 \label{ex}
 \end{figure}

For example, using this, $\eqref{totest}$ can instead be written as (derivatives are now w.r.t.\ $t$)
\be 
\label{w11-tr}
\ba W^{(1)}_1(t,\kappa)&=W^{(1)}_1(t,0)-(2\pi \ri)^{\frac{1}{2}}\kappa^{IJ} \left({1\over 2}{\partial_I} \omega_J+\partial_I F^{(1)}\omega_J\right)\\ 
& -{(2\pi \ri)^{\frac{1}{2}}\over 2}\kappa^{KM}\kappa^{NP}\partial_{K} \partial_{M} \partial_{N} F^{(0)} \omega_P~.\ea\ee
 We are interested in studying the $\kappa$ transformation of the the integrated open string amplitudes (\ref{wtoa}), which will be of the form 
\be
\label{a-k-exp}
A^{(g)}_h(t,\kappa,z_1,\ldots, z_h)=\sum_{m=0}^{3g-3+2h}\frac{\kappa^m}{m!} {\rd^m A^{(g)}_h\over \rd\kappa ^m}(t,0,z_1,\ldots, z_h).
\ee
We will denote, 
\be
\label{mf}
A^{(g)}_0=  F^{(g)}~.
\ee
In the following we will always use (\ref{intder1})-(\ref{intder4}) to express the results in terms of derivatives. As explained above the results of the $\kappa$-expansion can be represented graphically in terms of surfaces. We will use the following prescription to represent our result graphically:
 \begin{enumerate}
\item For each $A^{(g)}_h(t,0)$  we draw a Riemann surface   with $g$ holes and $h$ legs sticking out.
\item For each derivative $\partial_I$ acting on  $A^{(g)}_h(t,0)$ we draw a puncture on the Riemann surface.
\item For each element $\kappa^{IJ}$ we draw a propagator connecting the $I^{th}$ puncture to the $J^{th}$ puncture.
\end{enumerate}
By integrating (\ref{w11-tr}) we find, 
\be
\ba
A^{(1)}_1(t,\kappa,z)&=A^{(1)}_1(t,0,z)+\kappa^{IJ} \left({1\over 2}{\partial_I} \partial_{J} A^{(0)}_1(t,0,z)+\partial_JA^{(1)}_0(t,0)\partial_IA^{(0)}_1(t,0,z)\right)\\ 
& +{1\over 2}\kappa^{KM}\kappa^{NP}\partial_{K} \partial_{M} \partial_{N} A^{(0)}_0 (t,0)\partial_{P} A^{(0)}_1(t,0,z).
\ea
\ee
 \begin{figure}
\centerline{\includegraphics[width=0.9\linewidth,angle=0]{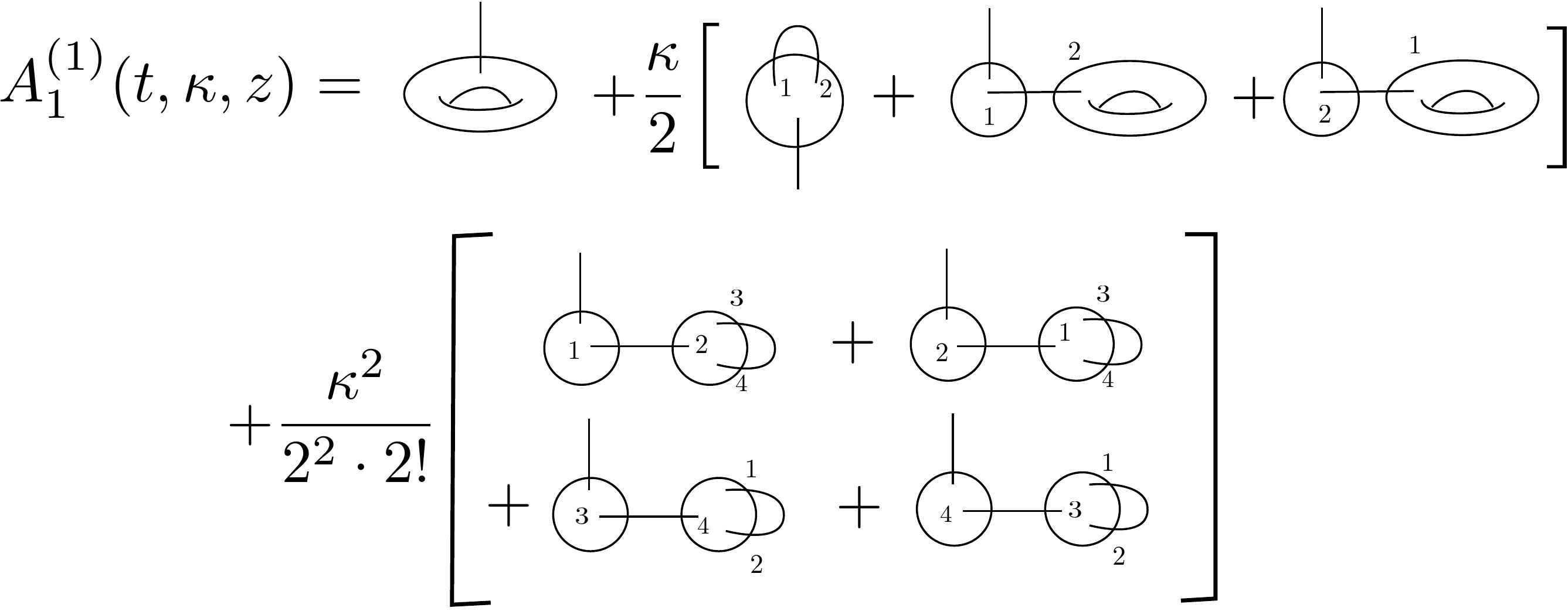} }
\caption{ Graphs that contribute to $A^{(1)}_1(t,\kappa,z)$ after iterating (\ref{variatW}).  }
 \label{grapha11}
 \end{figure} 
This can be represented graphically as in \figref{grapha11}.
In a similar way we obtain for the amplitude at genus zero and three boundaries, 
\be
\label{w03final}
\ba
&A^{(0)}_3(t,\kappa,z_1,z_2,z_3)= A^{(0)}_3(t,0,z_1,z_2,z_3) +3\sum\limits_{\sigma \in S_3}{1\over 3! }\kappa^{IJ}\partial_IA^{(0)}_2(t,0,z_{\sigma(1)},z_{\sigma(2)})\partial_JA^{(0)}_1(t,0,z_{\sigma(3)})
 \\ & \qquad + 3 \sum\limits_{\sigma \in S_3}{1\over 3! } \kappa^{IJ}\kappa^{KL} \partial_IA^{(0)}_1(t,0,z_{\sigma(1)})\partial_KA^{(0)}_1(t,0,z_{\sigma(2)})\partial_L\partial_JA^{(0)}_1 (t,0,z_{\sigma(3)})\\
 & \qquad + \kappa^{IJ}\kappa^{KL}\kappa^{MN}\partial_IA^{(0)}_1(t,0,z_1)\partial_KA^{(0)}_1(t,0,z_2)\partial_MA^{(0)}_1(t,0,z_3)\partial_J\partial_L\partial_NA^{(0)}_0(t,0).
 \ea 
 \ee
The graphs contributing to this result, up to order $\kappa$, are shown in  \figref{graphproof}.
For the amplitude at genus one and two boundaries,
 \be\label{w12final} \ba
&A^{(1)}_2(t,\kappa,z_1,z_2)\\
& \quad = A^{(1)}_2(t,0,z_1,z_2)+\kappa^{IJ}\left({1\over 2} \partial_I\partial_J A^{(0)}_2(t,0,z_1,z_2) +\partial_IA^{(1)}_0(t,0)\partial_J A^{(0)}_2(t,0,z_1,z_2) \right. \\ 
&\quad  \left. + 2\partial_IA^{(0)}_1(t,0,z_1)\partial_J A^{(1)}_1(t,0,z_2)  \right) +\kappa^{IJ}\kappa^{KL}\left( {1\over 2} \partial_IA^{(0)}_2(t,0,z_1,z_2)\partial_J \partial_L \partial_K A^{(0)}_0(t,0) \right.\\
& \quad\left.+ {1\over 2}  \partial_I \partial_K A^{(0)}_1(t,0,z_1) \partial_J\partial_L A^{(0)}_1 (t,0,z_2)+ {1\over 2}\partial_IA^{(0)}_1 (t,0,z_1) \partial_J \partial_K\partial_L A^{(0)}_1 (t,0,z_2)  \right.
 \\
 & \quad{1\over 2}\partial_IA^{(0)}_1 (t,0,z_2) \partial_J \partial_K\partial_L A^{(0)}_1 (t,0,z_1) + \partial_I \partial_L A^{(0)}_1(t,0,z_2) \partial_K A^{(0)}_1(t,0,z_1) \partial_J  A^{(1)}_0 (t,0)
 \\
& \quad \left. + \partial_I \partial_L A^{(0)}_1(t,0,z_1) \partial_K A^{(0)}_1(t,0,z_2) \partial_J  A^{(1)}_0 (t,0)+  \partial_I  A^{(0)}_1 (t,0,z_1)\partial_K A^{(0)}_1 (t,0,z_2)\partial_J \partial_L A^{(1)}_0  (t,0) \right)
\\
&\quad+ \kappa^{IJ} \kappa^{KL} \kappa^{MN}\left(  {1\over  2}   \partial_I  A^{(0)}_1(t,0,z_1) \partial_K A^{(0)}_1 (t,0,z_2) \partial_J  \partial_L \partial_M \partial_N A^{(0)}_0(t,0)+ \right.
\\ & \quad{1\over 2} \left(  \partial_I \partial_K A^{(0)}_1 (t,0,z_1)\partial_MA^{(0)}_1(t,0,z_2)+ \partial_I \partial_K A^{(0)}_1 (t,0,z_2)\partial_MA^{(0)}_1(t,0,z_1)\right)\partial_L \partial_J \partial_N A^{(0)}_0 (t,0)
\\ &\quad{1\over 2} \left( \partial_I \partial_KA^{(0)}_1(t,0,z_1)  \partial_L A^{(0)}_1(t,0,z_2)+ \partial_I \partial_KA^{(0)}_1(t,0,z_2)  \partial_L A^{(0)}_1(t,0,z_1)\right )\partial_M \partial_J \partial_N A^{(0)}_0(t,0)+
\\ 
&\quad\left. +\partial_I  A^{(0)}_1 (t,0,z_1)\partial_L A^{(0)}_1(t,0,z_2) \partial_K  \partial_J \partial_M A^{(0)}_0 (t,0)\partial_N   A^{(1)}_0(t,0)\right)\\
&\quad + \kappa^{IJ} \kappa^{KL} \kappa^{MN}\kappa^{PQ}\left( {1\over 2}   \partial_I  A^{(0)}_1(t,0,z_1) \partial_K A^{(0)}_1(t,0,z_2) \partial_L  \partial_J \partial_M A^{(0)}_0 (t,0)\partial_P  \partial_Q \partial_N  A^{(0)}_0(t,0)\right.
\\
&\quad\left.     + {1\over  2}   \partial_I  A^{(0)}_1 (t,0,z_1)\partial_K A^{(0)}_1(t,0,z_2) \partial_P  \partial_L \partial_M A^{(0)}_0 (t,0)\partial_J  \partial_Q \partial_N  A^{(0)}_0 (t,0)  \right).
 \ea\ee
As in \cite{abk,emo}, we would like to write these transformations in terms of an integral transform. We will first make an educated guess based on the above results, and then we will give a combinatorial proof in the next subsection. To proceed, we introduce the open string amplitudes to all genera and fixed number of boundaries, 
\be
\label{all-genera}
A_h (z_1,\ldots,z_h) = \sum_{g\ge 0} g_s^{2g-2+h} A_h^{(g)} (z_1,\ldots,z_h).
\ee
The above results for the $\kappa$-dependence of the open string amplitudes can be obtained from the following integral formulae, 
 \be \ba
A_1(t,\kappa,z)=&\int \rd\eta \, A_1(\eta, 0,z)\re^{-S(t,\eta,\kappa)g_s^{-2} + F(\eta,0) }\biggr|_{\text{connected}},\\
A_2(t,\kappa,z_1,z_2)=&\int \rd\eta \left(A_2( \eta,0,z_1,z_2) +A_{1}( \eta,0,z_1)A_{1}( \eta,0,z_2)\right) \re^{-S(t,\eta,\kappa)g_s^{-2} + F(\eta,0) }\biggr|_{\text{connected}}~,\\
A_3(t,\kappa,z_1,z_2,z_3)=&\int \rd \eta \bigl( A_3(\eta,0,z_1,z_2,z_3) +3  \sum\limits_{\sigma \in S_3}{1\over 3! }A_2(\eta,0,z_{\sigma(1)},z_{\sigma(2)}) A_1(\eta,0,z_{\sigma(3)}) \\ 
&  \, \,  \quad \quad + A_1(\eta,0,z_1)A_1(\eta,0,z_2)A_1(\eta,0,z_3)\bigr) \re^{-S(t,\eta,\kappa)g_s^{-2} + F(\eta,0) }\biggr|_{\text{connected}}.
 \ea
 \ee
In these equations, $S(t,\eta,\kappa)$ is given in (\ref{noyau}), and one performs the integrals by doing a saddle-point expansion at small $g_s$. As in \cite{emo}, the terms obtained when doing this expansion can be written in terms of the same diagrams that we considered before. 
One finds both connected and disconnected diagrams. To understand what connected means in this context, let us consider the following example:
 \be \ba & \kappa^{IJ}\kappa^{MN}\partial_I \partial_J A^{1}_{2}(t,0,z_1,z_2)\partial_M \partial_N A^{2}_{1}(t,0,z_3), \\
&  \kappa^{IJ}\kappa^{MN}\partial_I \partial_M A^{1}_{2}(t,0,z_1,z_2)\partial_N \partial_JA^{2}_1(t,0,z_3)
 .\ea\ee
  The diagrammatic representation of the above surfaces is given in  \figref{ex}.
The first one consists of two disconnected parts, while in the second one the two surfaces are linked together, i.e. they are connected.
 
This suggests the following expression for the $\kappa$-dependence, 
 \be\label{general}
 \ba
A_h(t,\kappa,z_H)=\int \rd\eta & \left(A_h( \eta,0,z_H)+\sum_{a=2}^{h}{1\over a!} \sum _{(L_1,\ldots,L_a)}A_{|L_1|}( \eta,0,z_{L_1}) \ldots A_{|L_a|}( \eta,0,z_{L_a}) \right)\, \\
& \times \re^{F(\eta,0)-S(t,\eta,\kappa)g_s^{-2}}\biggr|_{\text{connected}}
\ea\ee
where $L_i \subset H, L_i \neq \emptyset $ and $\sum_{i=1}^a L_i=H$. 
We can reorganise (\ref{general}) into a more elegant expression by including as well disconnected diagrams: 
\be \label{general2} 
\ba 
& \exp \left[\sum \limits_{h \geq 0}  {1 \over h!} \sum\limits_{{\boldsymbol \ell}} A_{{\boldsymbol \ell}}(t, \kappa)  \tr V^{\ell_1}\cdots\tr V^{\ell_h}\right] \\
  & = \int \rd\eta  \exp \left[ \sum \limits_{h \geq 0}  {1 \over h!} \sum \limits_{{\boldsymbol \ell}} A_{{\boldsymbol \ell}}(\eta, 0)  \tr V^{\ell_1}\cdots\tr V^{\ell_h}\right] \re^{-S(t,\eta,\kappa)g_s^{-2}}
  \ea \ee
where $A_{{\boldsymbol \ell}}$ is defined as in (\ref{a-ah}), but summed to all genera, 
\be 
\label{Aktkappa}
A_h (t,\kappa,z_1,\ldots, z_h)=  \sum \limits_{{\boldsymbol \ell}} A_{{\boldsymbol \ell}}(t, \kappa)  z_1^{\ell_1}\cdots z_h^{\ell_h}.
\ee
In writing these expressions, we have used the dictionary 
\be
\tr V^{\ell_1}\cdots\tr V^{\ell_h} \leftrightarrow{1\over h!}\sum\limits_{\sigma \in S_n}\prod \limits_{i=1}^h z_{\sigma(i)}^{\ell_i} .\ee
Notice that the quantity appearing in (\ref{general2}) is just the total open free energy $F(V)$. Therefore, (\ref{general2}) says that the $\kappa$-dependent open free energy is obtained from the original one by the same integral transform. 

We can now use this integral transform to obtain the modular transformation of the open plus closed string amplitudes, generalizing in this way the result of \cite{abk,emo} to the open sector. We will now put the formula (\ref{general2}) in the form presented in \cite{abk}. This will also take care of the transformation of 
$F^{(0)}$. Let us consider a symplectic transformation (\ref{simp}), and let us define the bilinear functional associated to $\Gamma$, 
\be
\label{abk-no}
\CS_\Gamma(x, \tilde{x}) = -{1 \over 2} (C^{-1} D)_{JK} x^{J} x^{K} + (C^{-1})_{JK}x^{J} {\tilde x}^K -{1\over 2}(AC^{-1})_{JK} {\tilde x}^J {\tilde x}^K. 
\ee 
This is the bilinear entering the integral transform of \cite{abk} for the closed string amplitudes,
\be 
\label{Zaganagic}
{ Z}({\tilde x})= \int \,\rd{x} \;\re^{-\CS_{\Gamma}(x,{\tilde x})g_s^{-2}}
\, {Z}({x}).
\ee
  Given $\tilde x$, let $x_{cl}$ be defined by
\be
{\tilde x}^{I} = C^{IJ}\partial_{J}F^0(x_{cl})+ {D^{I}}_{J} {x}^J_{cl},
\ee
which is the saddle-point for the integral transform (\ref{Zaganagic}). 
One has the following relationship, 
\be
\label{no-rel}
\CS_\Gamma(x,\tilde{x})=S(x_{cl}, x, -(\tau +C^{-1}D)^{-1})+\CS_{\Gamma}(x_{cl},\tilde{x})~,
\ee
where we have set $\eta=x$, $t=x_{cl}$. The first term in the r.h.s.\ in $\eqref{no-rel}$ leads to the integral kernel appearing in $\eqref{general2}$. 
The second term leads to a constant factor 
\be
\exp(g_s^{-2} \CS_{\Gamma}(x_{cl},\tilde{x}))
\ee
in front of the integral. This factor just gives the correct modular transformation of $F^{(0)}$, which is not incorporated in (\ref{general2}). If we now define, 
 \be \ba
 & A_h(\tilde{x}, z_1,\ldots, z_h):=A_h(x_{cl},-(\tau +C^{-1}D)^{-1},z_1,\ldots, z_h), \\
 & A_h(x,z_1,\ldots, z_h):=A_h(x,0,z_1,\ldots, z_h),
 \ea
\ee
where the quantities in the r.h.s.\ are defined in (\ref{all-genera}), and we modify (\ref{general2}) by using (\ref{abk-no}), we obtain the following formula for the modular transformation of the open string partition function, 
\be \label{final-mod} 
\ba 
& \exp \left[\sum \limits_{h \geq 0}  {1 \over h!} \sum\limits_{{\boldsymbol \ell}} A_{{\boldsymbol \ell}}(\tilde x)  \tr V^{\ell_1}\cdots\tr V^{\ell_h}\right] \\
  & = \int \rd x  \exp \left[ \sum \limits_{h \geq 0}  {1 \over h!} \sum \limits_{{\boldsymbol \ell}} A_{{\boldsymbol \ell}}(x)  \tr V^{\ell_1}\cdots\tr V^{\ell_h}\right] \re^{-\CS_\Gamma (x, \tilde x)g_s^{-2}}.
  \ea \ee
This generalizes (\ref{Zaganagic}) to the open sector. 

One can also use (\ref{general2}) to study the non-holomorphic dependence of the quantities $A_{\ell}(t, -(\tau-\overline \tau)^{-1})$. Since the non-holomorphic dependence in the r.h.s.\ of (\ref{general2}) is only due to the one appearing in $S(t, \eta, -(\tau-\overline \tau)^{-1})$, the open string partition function satisfies the 
same holomorphic anomaly equation as the closed string partition function. It is easy to show that one recovers, in particular, the holomorphic anomaly equation for open string amplitudes derived in \cite{emo}.

\subsection{A proof}

We now prove the relationship (\ref{general2}), i.e.\ we prove that the Feynman expansion of the integral on the r.h.s. 
generates the terms which are obtained by iterating (\ref{variatW}). Our proof is a generalization of the one for the closed string sector in \cite{emo}. 

The main idea is that we have the same kind of diagrams appearing in both sides of (\ref{general2}) and   we have to check  that each of them appears with the same multiplicity.  More precisely we will  show that the multiplicity factor appearing when we expand  the l.h.s.\  in $\kappa$ is the multiplicity factor which arises by applying Wick's theorem to the  r.h.s.\  of (\ref{general2}).
 
Let us first consider the l.h.s.\  and we start by looking at a single surface
 $A^{(g)}_h(t,\kappa,z_1, \ldots,z_h)$. By iterating (\ref{variatW}), 
the $\kappa$-expansion in (\ref{a-k-exp}) can be written as
\be
A^{(g)}_h(t,\kappa,z_1, \ldots,z_h)=\sum_{m=0}^{3g-3+2h} \sum_{I_1,\dots,I_{2m}} \kappa_{I_1,I_2} \dots \kappa_{I_{2m-1},I_{2m}} {1\over 2^m\, m!} \sum_{G_m^h} A_{G_m^h},
\ee
where $G_m^h$ is a connected, degenerate surface with $h$ legs sticking out and $m$ propagators connecting 
$2m$ points labelled by $1,\dots,2m$, in such a way that the point labelled by $2i-1$ is connected by a propagator 
to the point  labelled by $2i$. Each of the terms $A_{G_m^h}$ is of the form 
\footnote{\text{ The notation $\partial^{m_i}  A^{g_i}_{h_i}(t,0,z_1, \ldots,z_{h_i})$ 
means that there are $m_i$ derivatives acting on 
 $ A^{g_i}_{h_i}(t,0,z_1, \ldots,z_{h_i})$. }}
\be   
\prod_{i=1}^r  \partial^{m_i}  A^{(g_i)}_{h_i}(t,0,z_1, \ldots,z_{h_i}), 
\ee
 where %
 \be\label{selectionrule}
 \ba 
 \sum_i m_i&=2m, \\ 
 \sum_i h_i&=h, \\
 \sum_i (2g_i-2) +2m&=2g-2.
 \ea
 \ee
  Many $G_m^h$ graphs give the same contribution, hence it would be useful to count them only once and add a multiplicity factor. We can split the counting into two parts: first we compute the multiplicity factor due to the relabeling of external legs and then we compute the multiplicity due to the relabeling of the endpoints of the propagator. 
 \begin{itemize}
 \item  Let $\{h_1,\ldots,h_a\}=\{L_1,\ldots,L_1,L_2,\ldots,L_2,\ldots,L_a,\ldots,L_a\} $, where $L_i$ appear with multiplicity $n_i$, i.e.\  we  have $n_i$ surfaces with $L_i$ legs sticking out. The number of such terms is given by 
 \be
 {h! \over (L_1!)^{n_1}\cdots(L_a!)^{n_a}}.
 \ee
 \item For a fixed choice of $\{(L_1,n_1),\ldots,(L_a,n_a)\}$ we have an additional 
 factor $\mathcal{N}_{G_m^h}$, which counts the inequivalent ways of relabeling 
 the punctures at the endpoints of the propagators in such a way that $2i-1$ is linked to $2i$ for $i=1,\ldots,m$.
 \end{itemize}
 Hence each diagram appears with a multiplicity factor 
 \be 
 {h! \over (L_1!)^{n_1}\cdots(L_a!)^{n_a}} {\mathcal{N}_{G_m^h}\over 2^m m!}.
 \ee
 As  in \cite{emo} the factor $\mathcal{N}_{G_m^h} $ is related to the symmetry factor $s$ through
 \be\label{Ns}
 {\mathcal{N}_{G_m^h}\over 2^m m!}={1\over s}.
 \ee
The factor $1/s$ is the multiplicity  arising from Wick's theorem applied to
\be \int \rd\eta  \exp \left[ \sum \limits_{k \geq 0}  \sum \limits_{{\boldsymbol \ell}} A_{{\boldsymbol \ell}}(\eta, 0)  \tr V^{\ell_1}\cdots\tr V^{\ell_k}\right] \re^{-S(t,\eta,\kappa)g_s^{-2}},\ee
where we select only connected diagrams fulfilling $\eqref{selectionrule}$ and we fix the external legs. In particular, here we do not allow for permutation of the external legs.
 The relation (\ref{Ns}) follows from the definition of the symmetry factor. Indeed the symmetry factor counts the number of equivalent contractions which in our language is the number of equivalent ways of relabeling the punctures at the endpoints of the propagators. By standard combinatorics one can deduce the relation (\ref{Ns}).
It follows that 
\be \label{symfactor}  \mathcal{M}_{G_m^h}:={1 \over (L_1!)^{n_1}\cdots(L_a!)^{n_a}} {\mathcal{N}_{G_m^h}\over 2^m m!},\ee
is  the multiplicity factor for connected diagrams fulfilling $\eqref{selectionrule}$  arising from Wick's theorem applied to
\be \int \rd\eta  \exp \left[ \sum \limits_{k \geq 0} {1\over k!} \sum \limits_{{\boldsymbol \ell}} A_{{\boldsymbol \ell}}(\eta, 0)  \tr V^{\ell_1}\cdots\tr V^{\ell_k}\right] \re^{-S(t,\eta,\kappa)g_s^{-2}}, \ee
where we now allow for permutations of the external legs.
It follows that $A^{(g)}_{h}(t,\kappa,z_1, \ldots,z_{h})/h!$ is obtained by considering all connected diagrams fulfilling $\eqref{selectionrule}$ coming from
\be \int \rd\eta  \exp \left[ \sum \limits_{k \geq 0} {1\over k!}
 \sum \limits_{{\boldsymbol \ell}} A_{{\boldsymbol \ell}}(\eta, 0)  \tr V^{\ell_1}\cdots\tr V^{\ell_k}\right] \re^{-S(t,\eta,\kappa)g_s^{-2}}. \ee
From the Fourier transform point of view, imposing the selection rule $\eqref{selectionrule}$ is natural since it is equivalent to picking up only the terms 
which are proportionals to $g_s^{2g+h-2}$ in the genus expansion, with $h$ legs sticking out.  
As an example of this procedure, we show in \figref{graphproof} the graphs that contribute to $A^{(0)}_3 (t,\kappa,x,y,z)$ up to first order. We see that the multiplicity factor is precisely (\ref{symfactor}).

Let us consider the full term on the l.h.s.\ of $\eqref{general2}$. 
The argument of the exponential is made of products of connected surfaces. 
By expanding it we obtain a sum involving  terms of the form:  
\be \prod\limits_{i=1}^{s} \left(\mathcal{M}_{G_{m^{(i)}}^{h^{(i)}}}\prod\limits_{j=1}^{r^{(i)}} \partial^{m^{(i)}_j}A^{(g^{(i)}_j)}_{h^{(i)}_j}(t,0,z_1, \ldots,z_{h^{(i)}_j})\right)^{\tilde{n}_i} {1 \over \tilde{n}_1!\cdots\tilde{n}_s!},\ee
where each term 
\be \left(\mathcal{M}_{G_{m^{(i)}}^{h^{(i)}}}\prod\limits_{j=1}^{r^{(i)}} \partial^{m^{(i)}_j}A^{(g^{(i)}_j)}_{h^{(i)}_j}(t,0,z_1, \ldots,z_{h^{(i)}_j})\right)\ee
denotes a connected surface appearing with multiplicity $\tilde{n}_i$. 
Hence the total multiplicity factor is 
\be\prod \limits_{i=1}^s \left(\mathcal{M}_{G_{m^{(i)}}^{h^{(i)}}} \right)^{\tilde{n_i}}{1\over \tilde{n}_i!}. \ee
This is precisely the symmetry factor arising from Wick's theorem  applied to r.h.s.\ of $\eqref{general2}$. Indeed the multiplicity factor is inversely proportional to 
the equivalent ways of relabeling the punctures. This has two sources: 
\begin{enumerate}
\item  The equivalent ways of relabeling the puncture inside a given connected surface.
\item The equivalent ways of relabeling the puncture between  disconnected surfaces.
\end{enumerate} 
As explained above the first contribution is
\be \mathcal{M}_{G_{m^{(i)}}^{h^{(i)}}},\ee
while the second one is given by the overall factor
  \be{1 \over \tilde{n}_1!\cdots\tilde{n}_s!} .\ee

\sectiono{Application: the ABJM $1/2$ BPS Wilson loop}

As an application of the general result of this paper, we will now present expressions for the vevs of $1/2$ BPS Wilson loops 
of ABJM theory, in different representations. Since the results are obtained as an integral transform of topological vertex results, they are exact in the string coupling 
constant but perturbative in the exponentiated K\"ahler parameter. Therefore, they correspond to an expansion at large $N$ with $k$ fixed, which is the M-theory expansion of the 
Wilson loop amplitudes. 
 
 \subsection{ABJM theory and topological strings}
 
 The partition function and vevs of BPS Wilson loops in ABJM theory can be computed through localization (see \cite{mm-loc} for a review and a list of references). 
The result of this computation is a matrix integral \cite{kwy} which in turn can be related \cite{mp} 
to topological string theory in the CY manifold known as local $\IP^1 \times \IP^1$. This manifold has two K\"ahler moduli. In the large radius frame, 
they correspond to the sizes $T_1$, $T_2$ of the two $\IP^1$'s. ABJM theory is 
described by the ``slice"
\be
\label{ttt}
T=T_1=T_2. 
\ee
 The appropriate frame for the matrix model calculation in ABJM theory is the so-called orbifold frame of \cite{akmv}, and the strong 
coupling limit of ABJM theory corresponds to the large radius regime of topological 
string theory. One can then compute ABJM quantities at strong coupling by first computing the amplitudes 
in the large radius frame, and then performing a modular transformation. 

In the case of the partition function, one can use the integral transform formula of \cite{abk} to obtain the partition function of ABJM theory 
at strong coupling \cite{mp-fermi}. Let us review this in some detail. 
In the orbifold frame, the natural periods are 
\be
\lambda={N \over k}, 
\ee
which is the 't Hooft coupling of the gauge theory, and the dual period $\partial_\lambda F_0$, where $F_0$ is the genus zero free energy. In the large radius, 
the natural periods are $T$ (the diagonal K\"ahler modulus in (\ref{ttt})) and the derivative $\partial_T F_0^{\rm LR}$. There is only one effective class, labelled by an integer $d$, such that 
$\beta\cdot T= d T$. The closed string free energy in the large radius frame is given by 
\be
\label{LRfree}
 F^{\rm LR}(\lambda, g_s)=\frac{T^3}{6g_s^2}+\frac{T}{12}+A(g_s)+ \sum_{g\ge0}\sum_{d>0} N_{g,d} \, \re^{-dT}  g_s^{2g-2} ,
\ee
where $N_{g,d}$ are Gromov--Witten invariants in the local $\IP^1 \times \IP^1$ geometry and $A(g_s)$ is the contribution of constant maps. 
The topological string coupling constant is related to $k$ by\footnote{In \cite{mp,dmp,mp-fermi} one sets $g_s=2 \pi \ri/k$, but then the resulting topological string free energy at genus $g$ and large radius differs by a factor of $4^{g-1}$ from the standard one. The normalization used in this paper is more suited to comparisons with standard large radius results.}
\be
g_s= {4 \pi \ri \over k}.
\ee
There is a symplectic transformation relating the periods in the orbifold frame, to the periods in the large radius frame:
\begin{equation}
 \left(\begin{array}{c}
     \partial_{\tilde {\lambda}}\widetilde F_0 \\ \tilde{\lambda}
       \end{array}\right)=
\left(\begin{array}{cc}
 0 & 1 \\
 -1 & 2
\end{array}\right)
\left(\begin{array}{c}
    \partial_{\widetilde{T}} \widetilde{F}^{\rm LR}_0 \\  \widetilde{T}
       \end{array}\right)
\end{equation}  
where 
\be
\tilde{\lambda}={4\pi^2 \over c} \lambda, \quad \widetilde{T}={\pi \ri \over 2c} T, \qquad c^2=2\pi \ri,
\ee
and
\be
\label{shift}
\ba
\widetilde F_0&=F_0-\pi^3\ri\lambda,\\
 \widetilde{F}^{\rm LR}_{g}&=(-1)^{g-1}\left(F^{\rm LR}_{g}-\delta_{g,0}\frac{\pi^2T}{3}\right).
 \ea
\end{equation}
Then, according to (\ref{Zaganagic}), the total partition functions are related by the following formula:
\be
 \exp\left[F(\lambda)-\pi^3\ri\lambda/g_s^2\right] = \int \rd\widetilde{T}
\exp\left[-\widetilde{T}^2/g_s^2+\widetilde{T}\tilde{\lambda}/g_s^2+\widetilde{F}^{\rm LR}(\widetilde{T})\right].
\ee
Let us introduce a variable $\mu$ through 
\be
\label{Tmu}
T={4 \mu \over k}-\pi \ri. 
\ee
Then, one finds the expression
\be
\label{express}
\ba
Z_{\rm cl}(N, k)&=\re^{A(k)}
\int \rd \mu
\exp\left\{\frac{2\mu^3}{3 k\pi^2}-\mu N+\frac{k}{24}\,\mu+\frac{1}{3k}\,\mu+\CO\left(\re^{-\frac{4\mu}{k}}\right)\right\}\\ 
&=\re^{A(k)} \Ai\left[C^{-1/3}(N-B)\right]\left(1+{\CO}(\re^{-2 \pi {\sqrt{2 \lambda}}})\right),
\ea
\ee
where we used the following integral representation of the Airy function, 
\be
{\rm Ai}(z)={1\over 2 \pi \ri} \int_{\CC} \rd t\, \exp\left( {t^3 \over 3} -z t\right), 
\ee
and $\CC$ is a contour in the complex plane from $\re^{-\ri \pi/3}\infty$ to $\re^{\ri \pi/3}\infty$. In (\ref{express}), 
\be
C={2\over \pi^2 k},\qquad  B={k\over 24}+{1 \over 3k}. 
\ee
The result (\ref{express}) was first obtained in \cite{fhm} by studying the holomorphic anomaly equations. The function $A(k)$ has been studied in detail in \cite{honda}. 
\subsection{Wilson loops} 

It is possible to construct $1/2$ BPS Wilson loops in ABJM theory and to evaluate their vevs through localization \cite{dt}. 
These Wilson loops are labelled by Young tableaux $R$, and their vevs 
reduce to an average in the matrix model of \cite{kwy}. This has been recently tested in perturbation theory, to two loops, and for 
the fundamental representation, in \cite{bglp}. In \cite{mp} it was shown that the matrix model averages are topological 
open string amplitudes associated to an outer brane in local $\IP^1 \times \IP^1$, again in the so-called orbifold frame. 
We can now generalize to the open string sector the observation of \cite{mp-fermi} for computing the closed string partition function 
in the orbifold frame: we first evaluate the open string amplitudes in the large radius frame, 
and then perform an integral transform to obtain the result in the orbifold frame, which gives the Wilson loop vevs. 

\begin{table}
\begin{center}
\begin{tabular}{|| l l l l l l l ||}
\hline
$d$ & $0$ & $1$ & $2$ & $3$ & $4$ & $5$ \\ \hline
$g=0$ & $1$ & $2$ & $3$ & $10$ & $49$ & $288$ \\
$g=1$ & $0$ & $0$ & $0$ & $0$ & $8$ & $144$\\
$g=2$ & $0$ & $0$ & $0$ & $0$ & $0$ & $22$\\\hline
\end{tabular} \end{center}
\caption{The integer invariants $n_{g, d, (1)}$.}
\label{table1}
\end{table}
The open string amplitude for an outer brane in local $\IP^1 \times \IP^1$ can be computed by using for example the topological vertex \cite{akmv}. 
The topological vertex formalism computes directly the open string partition function $Z(V)$, and 
explicit results for the first $Z_R$ defined in (\ref{zr}) 
can be easily obtained. Equivalently, one can list the integer invariants $n_{g, d, \boldsymbol \ell}$ appearing in (\ref{multopen}). 
We find, for the fundamental representation, 
\be
\ba
\frac{Z_{\tableau{1}}}{Z_{\rm cl}}={1\over q-q^{-1}} \biggl[ &1+ 2 Q + 3Q^2 + 10 Q^3+ \left( 33+ 8 \left( q^2 + q^{-2} \right) \right) Q^4 \\ 
& + \left(132 +56\left(q^2 + q^{-2}\right)  + 22 \left(q^4 + q^{-4} \right) \right) Q^5 + \cdots\biggr]. \\
\ea
\ee
The integer invariants for ${\boldsymbol \ell} =(1)$ and ${\boldsymbol \ell}=(1,1), \, (2)$ are listed in tables \ref{table1} and \ref{table2}, respectively. 

\begin{table}
\begin{center}
 \qquad  \begin{tabular}{|| l l l l l l ||}
\hline
$d$ &   $1$ & $2$ & $3$ & $4$ & $5$ \\ \hline
$g=0$ &   $1$ & $2$ & $8$ & $48$ & $336$ \\
$g=1$ &  $0$ & $0$ & $0$ & $7$ & $148$\\
$g=2$ &  $0$ & $0$ & $0$ & $0$ & $20$\\\hline
\end{tabular} \qquad  
\begin{tabular}{||  l l l l l l ||}
\hline
$d$ &  $1$ & $2$ & $3$ & $4$ & $5$ \\ \hline
$g=0$  & $1$ & $2$ & $8$ & $36$ & $208$ \\
$g=1$ &  $0$ & $0$ & $0$ & $7$ & $112$\\
$g=2$ & $0$ &   $0$ & $0$ & $0$ & $20$\\\hline
\end{tabular} \end{center}
\caption{The integer invariants $n_{g, d, (1,1)}$ (left) and $n_{g, d, (2)}$ (right).}
 \label{table2}
\end{table}

When using our general result (\ref{final-mod}) we have to be careful with the open string mirror map. In the computation of large radius open string amplitudes with the topological vertex, we are implicitly using an open string modulus $\widetilde V$. Let $V$ be the open string modulus at  the orbifold point, appropriate for the matrix model of \cite{akmv,kwy}. Then, one has the relationship \cite{bkmp,mp,dmp}
\be 
\label{Mirrormap}\widetilde{V} =-Q^{-1/2} V=-\re^{T /2}V, 
\ee 
where we denoted
\be
Q=\re^{-T}. 
\ee
We will now test (\ref{final-mod}) for $h=1,2,3$ boundaries. 

\subsubsection{One boundary}

We will denote simply by $A_{\boldsymbol{\ell}}$ the open string amplitudes evaluated in the {\it orbifold} frame, 
which corresponds to ABJM vevs, and by $A_{\boldsymbol{\ell}}^{\rm LR}$ the amplitudes 
evaluated at large radius, computed for example by the topological vertex. To study the disk invariants, 
we specialize (\ref{final-mod}) for ${\boldsymbol{\ell}}=(l)$ and we pick only terms of the form $\tr \, V^l$. We find, 
\be \label{Disk}
\ba & Z_{\rm cl} (\lambda,k) \sum \limits_{l\geq 1} A_{l}(\lambda )\tr \, V ^l  \,\\
&=\sum \limits_{l\geq 1}\int \rd \mu \, A_{l}^{\rm LR}(T)\exp \left[ {2\mu ^3 \over 3 k \pi^2}-\mu N+\mu \left( {k\over 24}  +{1\over 3k} \right) + A(k)+ \CO\left(\re^{-{4\mu\over k}}\right)\right]\tr \, \widetilde{V}^l.
\ea\ee
We will now calculate the amplitudes by doing an expansion in $Q$, which corresponds to a worldsheet instanton expansion. 
From the results in table \ref{table1} we find, for $l=1$
\be
\label{LRQ}
A_1^{\rm LR}= {1\over 2 \ri \sin \left( {2 \pi \over k}\right)} \left(1+ 2 Q + 3 Q^2 + \cdots \right), 
\ee
and through the integral transform (\ref{Disk}) we obtain, for the leading contribution, 
 \be 
 \label{A1ORL} 
 A_1^{\rm lead} ={1\over 2 \sin (2 \pi /k)}\left( { \Ai \left(C^{-1/3}\left(N-{k\over 24}-{1\over 3k}-{2\over k}\right)\right)\over \Ai\left(C^{-1/3}\left(N-{k\over 24}-{1\over 3k}\right)\right) } \right).
 \ee
This reproduces the result derived in \cite{kmss} with Fermi gas techniques (up to an extra factor of $2$, due to the fact that we are expanding in $g_s={4\pi \ri \over k}$ rather than $g_s={2\pi \ri \over k}$ ). Our formalism makes it possible to calculate subleading corrections to this result coming from worldsheet instantons. In particular, the subleading order in $Q$ in (\ref{LRQ}) leads to the following exponentially small correction to (\ref{A1ORL}), 
\be \label{A1ORSL}  A_1^{\rm sl} =-\frac{1}{\sin(2 \pi /k)} W(-1)+{1\over 2 \sin ^3 (2 \pi /k)}W(-1)-{1\over 2 \sin ^3 (2 \pi /k)}W (-2)W(1),
\ee
where 
\be
W(n)=\frac{\Ai\left(C^{-1/3}\left(N-{k\over 24}-{1\over 3k}-{2 n\over k}\right)\right)}{\Ai\left(C^{-1/3}\left(N-{k\over 24}-\frac{1}{3k}\right)\right)}.
\ee
Higher order corrections can be computed straightforwardly from the topological vertex. The corrections in (\ref{A1ORSL}) and higher order should be interpreted, in the large $N$ type IIA superstring dual, as due to closed string worldsheet instantons attached to the disk instanton responsible for the leading order 
behavior (\ref{A1ORL}). 

Notice that the above formulae give {\it all genus results}: since the topological vertex expressions sum 
up the genus expansion, order by order in the degree, the expressions resulting from their 
integral transform sum up the genus expansion in the type IIA dual of ABJM theory, and can be therefore lifted to M-theory. 

One can also use the fact that, for $d=0$, the only non-zero integer invariant $n_{g,d, {\boldsymbol \ell}}$ occurs for $g=0$ and ${\boldsymbol \ell}=(1)$, to derive
\be
A_n^{\rm LR}= {1\over 2 n \ri \sin (2 n \pi / k)} +\cdots
\ee
at leading order in $Q$. This is due only to multicovering of the $n=1$ amplitude. Therefore we find
\be
  A_n^{\rm lead} ={1\over 2 n \sin (2 n \pi /k)}W(n), 
  \ee
which also agrees with the result in \cite{kmss} . 
 
We can now test the above all-genus results against explicit computations done directly in the orbifold frame. Following \cite{dmp}, we use the following genus expansion
\be
A_1=\sum_{g\ge 0}g_s ^{2g-1} A^{(g)}_{1}.
\ee
The quantities $A^{(g)}_1$ have been computed for $g=0$ \cite{mp} and $g=1$ \cite{dmp}, and they are naturally expressed in terms  of the parameter $\kappa$ introduced in \cite{mp} through 
\be
 \label{lamkap}
 \lambda(\kappa)={\kappa \over 8 \pi}   {~}_3F_2\left(\frac{1}{2},\frac{1}{2},\frac{1}{2};1,\frac{3}{2};-\frac{\kappa^2
   }{16}\right).
   \ee
This can be inverted, at strong coupling $\kappa \gg 1$, as
 \be
 \label{kappatolambda}
 \kappa= \re^{\pi \sqrt{2\hat{\lambda}}}\left(1+ \left(-2+{1\over \pi \sqrt{2\hat{\lambda}}}\right)\re^{-2\pi \sqrt{2\hat{\lambda}}} + \CO\left(\re^{-4\pi \sqrt{2\hat{\lambda}}}\right)\right), 
 \ee
 where 
 \be
 \hat{\lambda}=\lambda-{1\over 24}.
 \ee
 It has been shown in \cite{mp} that at genus zero the exact expression for the $1/2$ BPS Wilson loop vev is
 \be
 \label{WExact} A^{(0)}_1= \ri \kappa(\lambda), 
 \ee
 up to a factor of $1/2$ as mentioned after (\ref{A1ORL}). This is indeed reproduced by our formalism, since we find, from the genus expansion of (\ref{A1ORL}) and (\ref{A1ORSL}),  that
\be A^{(0)}_1= \ri  \, \re^{\sqrt{2\hat{\lambda}} \pi }+\ri 
\left({1\over \sqrt{2\hat{\lambda}}\pi}-2 \right) \re^{-\sqrt{2 \hat{\lambda}}\pi}  + \mathcal{O}\left(\re^{-3\pi \sqrt{2\hat{\lambda}}}\right),
\ee
which agrees with (\ref{WExact}) after using (\ref{kappatolambda}).
 
 We now look at the genus one results. From (\ref{A1ORL}) and (\ref{A1ORSL}) we have
 \be \label{A1ORLL} 
 \ba 
 A^{(1)}_1&= {3-4 \sqrt{2\hat{\lambda}}\pi + 4\hat{\lambda}\pi^2 \over 96 \ri \hat{\lambda} \pi^2}\re^{\pi \sqrt{2\hat{\lambda}}} 
 \\
 &-\frac{6+5  \sqrt{2\hat{\lambda}} \pi +4 \hat{\lambda} \pi ^2-20 \sqrt{2} \hat{\lambda}^{3/2} \pi ^3+16 \hat{\lambda}^2 \pi ^4}{192 \ri \hat{\lambda}^2 \pi ^4} \re^{-\pi \sqrt{2\hat{\lambda}}} +\mathcal{O}\left(\re^{-3\pi \sqrt{2\hat{\lambda}}}\right).
 \ea \ee
This can be compared with the expression obtained from the $W_1^{(1)}(p)$ computed in \cite{dmp}. We find perfect agreement. 

\subsubsection{Two boundaries}

 Let us now look at the terms proportional to $\left( \tr V \right)^2$ in (\ref{general2}):
\be \label{Annulus}\ba &  Z_{\rm cl} (\lambda, k)\left(A_{1,1}(\lambda)+ A_{1}(\lambda) A_{1}(\lambda) \right) \left( \tr V \right)^2 \, \\
&=\int \rd \mu \left(A_{1,1}^{\rm LR}(T)+A_{1}^{\rm LR}(T)A_{1}^{\rm LR}(T) \right) \\
&\qquad \qquad \times
\exp \left[ {2\mu ^3 \over 3 k \pi^2}-\mu N+ \mu \left( {k\over 24}  +{1\over 3k} \right) +A(k)+ \CO\left(\re^{-{4\mu\over k}}\right)\right]\left( \tr \widetilde V \right)^2.\ea\ee
The integrand in the second line can be computed by using the topological vertex, or equivalently the results for the integer invariants in table \ref{table2}. 
As before, we first look at the leading order in the worldsheet instanton expansion, and we find, after the integral transform, 
\be 
\ba
\label{A11} 
&A^{\rm lead}_{1,1}(\lambda)\\
&={1\over 4 \sin ^2 ({2\pi \over k })} \left[{ \Ai(C^{-1/3}(N-{k\over 24}-{1\over 3k}-{4\over k}))\over \Ai\left(C^{-1/3}\left(N-{k\over 24}-{1\over 3k}\right)\right) }-\left( { \Ai\left(C^{-1/3}\left(N-{k\over 24}-{1\over 3k}-{2\over k}\right)\right)\over \Ai\left(C^{-1/3}\left(N-{k\over 24}-{1\over 3k}\right)\right) } \right)^2 \right].
\ea
\ee
Similarly, the subleading term is given by
\be \label{A11ORSL} 
\ba  A^{\rm sl}_{1,1}=&\left({1\over 2}+{1\over 4 \sin^4(2\pi/k)}-{1 \over \sin^2(2\pi/k) } \right)W(0) +{1\over 2 \sin^4(2 \pi/k) }W(-2)W(1)W(1)\\
&+ \left({1\over \sin^2(2 \pi/k)}-{1\over 2 \sin^4(2 \pi/k) }\right)W(1)W(-1)-{1\over 4 \sin^4(2 \pi/k) }W(2)W(-2).
 \ea \ee
The annulus amplitude $\eqref{A11}$ has a genus expansion given by
\be
A_{1,1}=\sum\limits_{g \ge 0}{g_{s}^{2g}A^{(g)}_{1,1}}~.
\ee
We find, at genus zero, 
 \be \label{explead11} 
 A^{(0)}_{1,1}=-\frac{\re^{2\sqrt{2}\pi \sqrt{\hat{\lambda}}}}{8\sqrt{2}\pi\sqrt{\hat{\lambda}}} + \frac{\sqrt{2}-8 \sqrt{2} \hat{\lambda} \pi ^2+32 \hat{\lambda}^{3/2} \pi ^3}{32 \hat{\lambda}^{3/2} \pi ^3} + \mathcal{O}\left( \re^{-2\pi \sqrt{2\hat{\lambda}}} \right).
 \ee 
We can compute this quantity by using Akemann's expression for the annulus correlator $W^{(0)}_2(p,q)$ \cite{akemann}, which has an expansion
\beq
W(p,q)=\sum\limits_{k,l\geq 1}{kl A^{(0)}_{k,l} p^{-k-1}q^{-l-1}}~.
\eeq{}
The explicit expression for $W(p,q)$ is given in terms of data of the spectral curve of the ABJM matrix model, 
\beq
\begin{split}
W(p,q)=&\frac{1}{4(p-q)^2}\left(\sqrt{\frac{(p-x_1)(p-x_2)(q-x_4)(q-x_3)}{(p-x_4)(p-x_3)(q-x_1)(q-x_4)}}+\sqrt{\frac{(p-x_4)(p-x_3)(q-x_1)(q-x_2)}{(p-x_1)(p-x_2)(q-x_4)(q-x_3)}}\right)\\
&+\frac{(x_1-x_3)(x_4-x_2)}{4\sqrt{\sigma(p)\sigma(q)}}\frac{E(k)}{K(k)}-\frac{1}{2(p-q)^2}. 
\end{split}
\eeq{}
Here, $x_i$ are the branch points of the spectral curve, which is elliptic, $k$ is an appropriate elliptic modulus, and $E(k)$, $K(k)$ are elliptic integrals. Explicit 
expressions for all these quantities can be found in \cite{dmp}, section 8.1, or \cite{kmss}, section 2. The result obtained in this way matches with (\ref{explead11}). 
 
 \subsubsection{Three boundaries}
We finally discuss very briefly a simple check for the amplitude with three boundaries. 
Using the integral transform, we find, at leading order in the worldsheet instanton expansion, 
\beq
A_{1,1,1} = \left({\frac{1}{2\sin\left(\frac{2\pi}{k}\right)}}\right)^3\left(W(3)-3(W(2)-W(1)^2)W(1)-W(1)^3\right) +\cdots.
\eeq{3pt}
This has a genus expansion
\beq
A_{1,1,1} =\sum_{g\ge 0}g_s^{2g+1}A^{(g)}_{1,1,1}.
\eeq{}
At genus zero we find, from (\ref{3pt}), 
\be \label{A111expk} 
A^{(0)}_{1,1,1}=\frac{\ri}{64}\frac{(1-3\pi \sqrt{2\hat{\lambda}})}{(\pi \sqrt{2\hat{\lambda}})^3}\re^{3\pi \sqrt{2\hat{\lambda}}}+\CO\left(\re^{\pi \sqrt{2\hat{\lambda}}}\right).
\ee
We can compare this with the explicit expression extracted from the $W^{(0)}_3(p,q,r)$ presented in for example \cite{bkmp} (and applied to the spectral curve of the ABJM matrix model). The result at leading order matches again with (\ref{A111expk}).

\sectiono{Conclusions and open problems}

In this paper, motivated by the results in \cite{kmss}, we have generalized the results of \cite{emo} to the open sector, and showed that the 
intricate combinatorics of the $\kappa$-dependence in \cite{eo} can be simply summarized by the 
statement that the total open string partition function is a wavefunction. 
Notice that in our derivation the open moduli played no significant role, since the open mirror map simply involves multiplying them by an overall factor, and the only source 
of non-holomorphic dependence is the factor $\kappa$ already present in the closed sector. Cleary, it would be very interesting to generalize this wavefunction behavior under modular transformations to the case of compact CY manifolds. This can be 
probably worked out as a consequence of \cite{nw}. 

A nice application of our general result is the computation of vevs of $1/2$ BPS Wilson loops in ABJM theory. 
These are simply given by integral transforms of open string amplitudes in the 
large radius frame. The expressions for these amplitudes in terms of integer invariants, which 
include all genera but are perturbative in the exponentiated K\"ahler coupling, correspond 
precisely, after the integral transform, to the M-theory expansion of Wilson loop vevs. By using the large $N$ dual, we obtain in this way a genus resummation 
in type IIA superstring theory.  

One important open problem which we have not addressed here is the computation of membrane instanton corrections to the 
Wilson loop vevs of ABJM theory. These corrections are known to be present in the free energy on the three-sphere \cite{dmp-np,mp-fermi} and they can be computed 
within the Fermi gas formalism \cite{mp-fermi,hmo1,cm,hmo2}. It would be very interesting to know 
whether they are present in the case of Wilson loops, and if so, what is their value. 

In this respect there is however one interesting 
difference between the vevs of $1/2$ BPS Wilson loops and the 
free energy. For the free energy, the contribution of worldsheet instantons is singular for all integer values of $k$. This singular 
behavior is not physical (the original matrix integral is well-defined for all $k>0$), and as shown in \cite{hmo1} 
these singularities are cancelled by membrane instantons. In the case of vevs of $1/2$ BPS Wilson loops with winding numbers $\boldsymbol{\ell}$, there are singularities for the values of $k$ which divide $2\ell_i$, for all $i=1, \cdots, h$ (this follows from the integrality structure (\ref{multopen})). For example, for the winding $\boldsymbol{\ell}=(1)$, the vev is singular for $k=1,2$. However, as pointed out in \cite{kmss}, these singularities are physical, since the matrix integral computing the vevs actually diverges for these values. 
Therefore, the contribution of membrane instantons 
is not required to cancel out singularities, as in the case of the free energy, and it might be zero. It would be interesting to further explore this issue. 

\section*{Acknowledgements}
We would like to thank Albrecht Klemm for discussions. This work is supported by the Fonds National Suisse,
subsidy 200020-137523. 

\appendix
\section{A check of modular transformation properties}
As explained in section (\ref{sec:symptransf}) the $\kappa$-dependence in the quantities $W^{(g)}_k$ encodes the information about how it transforms under a modular transformation. In this appendix we perform an explicit check of this statement for a simple case of a local Calabi--Yau manifold.
A particular class of manifolds considered in this paper are those having  an underlying algebraic curve which is given in "exponentiated" variables by
\beq
y^2(x)=M^2(x)\sigma(x),
\eeq{}
where 
\beq
\sigma(x)=\prod\limits_{i=1}^{2s}{(x-x_i)}
\eeq{}
and $M(x)$ is some transcendental function. Let us consider the simple case $s=2$ and $M(x)=1$. In this case we have a curve of genus one, i.e. an elliptic curve. The cuts defining the $\underline{\CA}$ and $\underline{\CB}$-cycles are given by
\beq
\underline{\CA}_1=(x_1,x_2), \quad \underline{\CA}_2=(x_3,x_4), \quad \underline{\CB}_1=(x_2,x_3) ~.
\eeq{}
We will consider an $S$-duality transformation defined by 
\beq
A=D=0, \quad B=-C=1
\eeq{}
in (\ref{Cycletransform}). According to $\eqref{mod-trans}$ the quantity $W^{(1)}_1(p)$ will transform as $\eqref{totest}$ with $\kappa=-\tau^{-1}$. We can test this by using explicit expressions in terms of the end points of the cuts for the various quantities appearing in $\eqref{totest}$. 
\\
\\
We have \cite{akemann,eynard} 

\begin{align}
W^{(0)}_3(p_1,p_2,p_3)&=\frac{1}{8}\sum\limits_{i=1}^{4}{M^2(x_i)\sigma'(x_i)\chi^{(1)}_i(p_1)\chi^{(1)}_i(p_2)\chi^{(1)}_i(p_3)},\label{w03akemann}\\
W_1^{(1)}(p)&={1\over 16} \sum_{i=1}^{4} \chi^{(2)}_i(p) +{1\over 8} \sum_{i=1}^{4} \biggl( 2{\alpha_i 
} -\sum_{j\not=i} {1\over x_i -x_j} \biggr) \chi_i^{(1)}(p), \label{W11akemann}
\end{align}
where
\be\ba
\chi^{(1)}_i(p)&=\frac{1}{M(x_i)\sqrt{\sigma(p)}}\left(\frac{1}{p-x_i}+\alpha_i\right),\\
\chi^{(2)}_i(p)&= -{M'(x_i)\over M(x_i)} \chi^{(1)}_i(p)+  {1\over M(x_i){\sqrt {\sigma (p)}}} {1\over (p-x_i)^2} -{1\over 3 } {1\over M(x_i){\sqrt {\sigma (p)}}} \sum_{j\not= i} {\alpha_j -\alpha_i \over 
x_j -x_i},
\ea\ee
and the  $\alpha_i$ are given by 
\be \label{alpha}
\ba
\alpha_1&={1\over(x_1-x_2)}\left[1-{(x_4-x_2)\over(x_4-x_1)}{E(k)\over K(k)}
\right],\cr
\alpha_2&={1\over(x_2-x_1)}\left[1-{(x_3-x_1)\over(x_3-x_2)}{E(k)\over K(k)}
\right],\cr
\alpha_3&={1\over(x_3-x_4)}\left[1-{(x_4-x_2)\over(x_3-x_2)}{E(k)\over K(k)}
\right],\cr
\alpha_4&={1\over(x_4-x_3)}\left[1-{(x_3-x_1)\over(x_4-x_1)}{E(k)\over K(k)}
\right]. 
\ea
\ee
The $S$-duality transformation for $W^{(1)}_1(p)$ is given by exchanging the roots
\beq
x_1\leftrightarrow x_3
\eeq{}
in $\eqref{W11akemann}$. Doing this we find after a little work that 
\be\ba
W^{(1)}_1(p)\rightarrow W^{(1)}_1(p) +{1\over \sqrt{\sigma (p)}}\sum_{i=1}^4F(p,x_i, \tau),\ea \ee
where
\be \label{F} \ba F(p,x_i, \tau) & ={1\over 4}{K_i \over (p-\tilde{x}_i)} -\sum_{j\neq i}\left({1\over 48} { K_j-K_i \over \tilde{x}_j-\tilde{x}_i} +{1\over 8}{ K_i\over \tilde{x}_i-\tilde{x}_j} \right)+{1\over 4}(2K_i+2K_iR_i+K_i^2) ,\ea\ee
with 
\be\label{xtildex}
\ba
K_1&=-{1\over(x_3-x_2)}\left[{(x_4-x_2)\over(x_4-x_3)}{\ri\pi\over 2 K(k)^2 \tau }
\right], \qquad R_1&=-{1\over(x_3-x_2)}\left[{(x_4-x_2)\over(x_4-x_3)}(1-{E(k)\over K(k)})
\right], \cr
K_2&=-{1\over(x_2-x_3)}\left[{(x_1-x_3)\over(x_1-x_2)}{\ri\pi\over 2 K(k)^2 \tau }
\right],\qquad R_2&=-{1\over(x_2-x_3)}\left[{(x_1-x_3)\over(x_1-x_2)}(1-{E(k)\over K(k)})
\right],\cr
K_3&=-{1\over(x_1-x_4)}\left[{(x_4-x_2)\over(x_1-x_2)}{\ri\pi\over 2 K(k)^2 \tau }
\right],\qquad R_3&=-{1\over(x_1-x_4)}\left[{(x_4-x_2)\over(x_1-x_2)}(1-{E(k)\over K(k)})
\right],\cr
K_4&=-{1\over(x_4-x_1)}\left[{(x_1-x_3)\over(x_4-x_3)}{\ri\pi\over 2 K(k)^2 \tau }
\right],  \qquad R_4&=-{1\over(x_4-x_1)}\left[{(x_1-x_3)\over(x_4-x_3)}(1-{E(k)\over K(k)})
\right], \\
&\tilde{x}_i=\left\{ \begin{array}{l l}
x_i \quad \text{if} \quad i=2,4 \\
x_3 \quad \text{if} \quad i=1 \\
x_1 \quad \text{if} \quad i=3
\end{array}\right\}
\ea
\ee
and $\tau$ is the standard elliptic modulus
\beq
\tau=\ri\frac{K(k')}{K(k)}
\eeq{}
where 
\beq
k'^2=1-k^2~.
\eeq{}
Using the same approach as in for example \cite{bkmp} we can compute the integrals around the $\CB$-cycles of the quantities $W^{(0)}_2(p_1,p_2)$, $W^{(1)}_1(p)$ and $W^{(0)}_3(p_1,p_2,p_3)$, which are needed on the r.h.s\ of $\eqref{totest}$, in terms of elliptic functions. Without showing the full computation let us give a few important ingredients. For $W^{(0)}_2(p_1,p_2)$ we have \cite{bde}, 
\be
\oint\limits_{\mathcal{B}}W^{(0)}_2=2\pi \ri\omega=-{\pi \over 2 K(k)}\frac{\sqrt{(x_1-x_3)(x_2-x_4)}}{ \sqrt{\sigma(p)}}.\\
\ee
For the integrals of the quantities $\chi^{(1)}_i(p)$ in $\eqref{w03akemann}$ we find
\be \ba
\oint\limits_{\mathcal{B}}{\chi^{(1)}_1}&=-\frac{4}{\sqrt{(x_1-x_3)(x_2-x_4)}}{\pi \over 2}{1\over K(k)}{x_2-x_4 \over (x_1-x_2)(x_1-x_4)},\\
\oint\limits_{\mathcal{B}}{\chi^{(1)}_2}&=-\frac{4}{\sqrt{(x_1-x_3)(x_2-x_4)}}{\pi \over 2}{1\over K(k)}{x_3-x_1 \over (x_1-x_2)(x_2-x_3)},\\
\oint\limits_{\mathcal{B}}{\chi^{(1)}_3}&=-\frac{4}{\sqrt{(x_1-x_3)(x_2-x_4)}}{\pi \over 2}{1\over K(k)}{x_4-x_2 \over (x_3-x_4)(x_3-x_2)},\\
\oint\limits_{\mathcal{B}}{\chi^{(1)}_4}&=-\frac{4}{\sqrt{(x_1-x_3)(x_2-x_4)}}{\pi \over 2}{1\over K(k)}{x_1-x_3 \over (x_3-x_4)(x_4-x_1)}.\\
\ea\ee
In order to compute $  \oint\limits_{\mathcal{B}}W^{(1)}_1  $
it is useful to observe that 
\be\label{D1deriv}
\int\limits_{x_2}^{x_3} {\frac{\rd x}{(x_i-x)^2 \sqrt{|\sigma(x)|}}}=K(k') {1 \over 3}\sum_{j \neq i}{\tilde{\alpha}_j-\tilde{\alpha}_i\over \tilde{x}_j-\tilde{x}_i}.
\ee
 The quantities $\tilde{\alpha}_i$ and $\tilde{x}_i$ in (\ref{D1deriv})  are obtained from $\alpha_i$ and $x_i$ by everywhere exchanging the indices $1$ and $3$, as in $\eqref{xtildex}$.
\\
\\
After computing the relevant integrals and putting everything together we find that indeed 
\beq
\begin{split}
&-\frac{1}{2\pi \ri\tau} \left({1\over 2} \oint\limits_{\mathcal{B}}
\oint\limits_{\mathcal{B}}{W^{(0)}_3} +\oint\limits_{\mathcal{B}}W^{(1)}_1\oint\limits_{\mathcal{B}}{W^{(0)}_2}\right)+\frac{1}{(2\pi \ri\tau)^2}\left({1\over 2} \oint\limits_{\mathcal{B}}\oint\limits_{\mathcal{B}} \oint\limits_{\mathcal{B}}{W^{(0)}_3} \oint\limits_{\mathcal{B}}W^{(0)}_2 \right) \\
&={1\over \sqrt{\sigma (p)}}\sum_{i=1}^4F(p,x_i, \tau)~,
\end{split}
\eeq{}
where $F(p,x_i,\tau)$ is defined in $\eqref{F}$. This is what we wanted to show.

\end{document}